\documentclass[reprint,aps,amsmath,amssymb,floatfix,longbibliography]{revtex4-2}

\usepackage{xcolor}
\usepackage{graphicx}

\usepackage{bm}
\usepackage{dcolumn}

\usepackage{amsmath}
\newcommand\norm[1]{\left\lVert#1\right\rVert}

\usepackage[
colorlinks=true,
urlcolor=blue,
citecolor=blue,
linkcolor=blue,
hyperfootnotes=false]{hyperref}

\begin{document}
\title{Dynamics of quantum spin-nematics: Comparisons with canted antiferromagnets}
\author{Tsutomu Momoi}
\affiliation{Condensed Matter Theory Laboratory, RIKEN,
Wako, Saitama 351-0198, Japan}
\affiliation{RIKEN Center for Emergent Matter Science, Wako, Saitama, 351-0198, Japan}

\date{\today}

\begin{abstract}
The identification of spin-nematic states is challenging due to the absence of Bragg peaks.
However, the study of dynamical physical quantities provides a promising avenue for characterizing these states.
In this study, we investigate the dynamical properties of spin-nematic states in
three-dimensional quantum spin systems in a magnetic field, using
a two-component boson theory that incorporates magnons and bi-magnons.
Our particular focus lies on the dynamical spin structure factor at zero temperature and the nuclear magnetic resonance (NMR) relaxation rate at finite temperatures.
Our findings reveal that the dynamical structure factor does not exhibit any diverging singularity across momentum and frequency while providing valuable information about the form factor of bi-magnon states and the underlying structure of spin-nematic order.
Furthermore, we find a temperature dependence in the NMR relaxation rate proportional to $T^3$ at low temperatures, similar to canted antiferromagnets. A clear distinction arises as there is no critical divergence of the NMR relaxation rate at the spin-nematic transition temperature.
Our theoretical framework provides a comprehensive understanding of the excitation spectrum and the dynamical properties
of spin-nematic states,
covering arbitrary spin values $S$ and encompassing site and bond nematic orders.
Additionally, we apply the same methodology to analyze these dynamical quantities
in a canted antiferromagnetic state and compare the results with those in the spin-nematic states.
\end{abstract}
\maketitle

\section{Introduction}
Exploring hidden order phases in condensed matter is challenging due to their elusive nature in conventional static measurements.
An intriguing hidden order is a spin-nematic order in spin systems, characterized by the absence of a magnetic Bragg peak
and the presence of a broken partial spin rotation symmetry resulting from spin quadrupolar order~\cite{AndreevG1984}.
The emergence of spin-nematic order arises from the Bose-Einstein condensation of
bound magnon pairs~\cite{ShannonMS}.
Theoretical investigations have predicted the appearance of spin-nematic order in various systems,
including spin-$S$ bilinear-biquadratic systems~\cite{Matveev1973,Papanicolaou1984,Tsunetsugu2006,LauchliMP2006},
frustrated
ferromagnets~\cite{Vekua2007,HikiharaKMF,Sudan2009,ZhitomirskyT2010,SatoHM2013,ShannonMS,Jiang2023,MomoiSK2012,UedaM2013,Janson2016},
and quantum dimer systems~\cite{MomoiT,WangB,Yokoyama2018,HikiharaSU4}.
Importantly, the symmetry of magnon pairing varies across systems, with
$S$-wave symmetric magnon pairs leading to quadrupolar order on each lattice site and magnon pairs with
other symmetries, such as $d$-wave symmetry, leading to bond nematic order with quadrupolar order on bonds.
Subsequent to these predictions,
there has been a growing interest in experimental investigations aimed at
verifying the existence of spin-nematic phases~\cite{Buttge2014,Nawa2017,Orlova2017,Grafe2017,Skoulatos2019,Yoshida2017,Kohama,Kimura,Imajo}.
However, conclusive experimental confirmation requires further insights into
the intrinsic characteristics of these phases.

Detecting spin-nematic phases is challenging,
but dynamical quantities show promise as valuable tools for identification.
In one-dimensional spin-nematic Tomonaga-Luttinger liquids, the temperature dependence of the
nuclear-magnetic-resonance (NMR)
relaxation rate $1/T_1$ shows a slower decay compared to conventional one-dimensional antiferromagnets~\cite{SatoMF},
providing a means to detect spin-nematic liquids (see, for example, Ref.~\cite{Grafe2017}).
However, in three-dimensional spin-nematic ordered phases, different methods have yielded
inconsistent results regarding the temperature dependence of the NMR relaxation rate at low temperatures~\cite{ShindouYM2013,SmeraldS2016}, leading to a lack of comprehensive understanding.
In a $1/N$ expansion, the temperature dependence of $1/T_1$ was estimated as $T^5$~\cite{ShindouYM2013},
whereas a field-theoretical approach concluded $T^7$~\cite{SmeraldS2016}.
Both results deviate from the temperature dependence of $T^3$ for the conventional antiferromagnets derived by Moriya~\cite{Moriya1956A,Moriya1956B,Jaccarino1965}.
Another distinct difference from antiferromagnets was argued to be the absence of a critical divergence in $1/T_1$ at the transition temperature of the spin-nematic phase~\cite{SmeraldS2016}.

In this paper, we investigated the dynamical properties of
spin-nematic phases in three-dimensional systems in a magnetic field to address this confusing issue.
To accomplish this, we developed a methodology for describing spin-nematic states using a boson representation,
which enables us to
use the established standard interacting boson theory.
Expanding on our previous study~\cite{Imajo},
where we employed a two-component boson theory including both magnons and bi-magnons
to investigate the thermodynamic properties
of the spin-gap phase near a spin-nematic phase,
the present work extends this methodology.
Specifically, we investigated the dynamical quantities in the spin-nematic and paramagnetic phases,
considering the influence of interactions and bi-magnon condensation.
Our approach incorporates the structure of spin-nematic order, as captured by the
form factor of bound magnon pairs, enabling the analysis of various types of spin-nematic order,
including site-nematic and bond-nematic orders, within a unified theoretical framework.
Consequently, this method enables a comprehensive study of the spin-nematic phases.

In particular, we calculated the dynamical spin structure factor at $T=0$ and
the NMR relaxation rate at finite temperatures.
We find that the dynamical structure factor shows no diverging singularity at any momentum and frequency,
consistent with previous studies on specific spin models~\cite{Tsunetsugu2006,LauchliMP2006,ShindouYM2013,SmeraldUedaShannon2015}.
Our results uncover that the dynamical spin structure factor carries information about the form factor of bi-magnon states, providing insights into the underlying structure of spin-nematic order.
Furthermore, in contrast to previous studies, we observe that the NMR relaxation rate in the spin-nematic phase exhibits a temperature dependence proportional to $T^3$, similar to canted antiferromagnets.
A notable distinction from antiferromagnets is the absence of a critical divergence in the NMR relaxation rate at the spin-nematic transition temperature, consistent with a prior field-theoretical study~\cite{SmeraldS2016}.
Among these results, explicit dependence on pairing symmetry appears only in the intensity of the one-magnon mode
in the dynamical structure factor.
The remaining results constitute universal features of spin-nematic phases that do not explicitly rely on pairing symmetry.
Additionally, using the same methodology, we reanalyzed these dynamical quantities in a canted antiferromagnetic phase,
facilitating comparisons between spin-nematic and antiferromagnetic systems.

This paper is organized as follows: Section \ref{sec:BEC_bimagnon} details the formulation
for describing spin-nematic phases
in a magnetic field. Section \ref{sec:dynam_SN} presents the results for the dynamical quantities
of spin-nematic phases.
Section \ref{sec:dimer} describes an application of our results to the spin-nematic phase
in the low magnetic field regime of spin-dimer systems.
Section \ref{sec:BEC_magnon} focuses on the analyses of canted antiferromagnets.
Finally, we summarize our main findings in Sec.\ \ref{sec:summary}, along with discussions.

\section{BEC of bi-magnons}\label{sec:BEC_bimagnon}
When magnetic excitations above a spin-gapped ground state form stable two-particle
bound states with lower energy than two scattering excited states,
these bound states close the gap as the magnetic field varies.
Subsequently, at low temperatures, a macroscopic number of magnetic bound particles
are induced by the field, interacting with each other, which leads to a spin-nematic
phase~\cite{ShannonMS}.
In this section, we introduce an interacting boson theory to describe the spin-nematic phase in a magnetic field.
In the Supplemental Materials of our previous paper~\cite{Imajo},
we discussed the characteristics of the spin-gapped phase adjacent to the spin-nematic phase.
Extending the previous analysis, we investigate the effects of interactions and
Bose-Einstein condensation (BEC) of bi-magnons.

\subsection{Interacting boson theory for spin nematics}

In the following theory, we treat the spin-nematic phases near saturation in spin-$S$ systems
on a three-dimensional lattice $\Lambda$, where $S$ can be arbitrary.
One example of such a system is a layered square-lattice frustrated ferromagnet, where spins on each layer
couple with the ferromagnetic 1st-neighbor and antiferromagnetic 2nd-neighbor interactions~\cite{ShannonMS,Jiang2023}.
These arguments are equally applicable to the spin-nematic phase observed in low fields in frustrated
spin-dimer systems,
including the two-dimensional Shastry-Sutherland model~\cite{MomoiT,WangB,Imajo,fogh2023},
as discussed in Sec.\ \ref{sec:dimer}.

\subsubsection{Boson mapping}

To describe dilute magnetic excitations on top of the fully polarized state,
we set up Fock spaces for both single magnon excitations ($S^z=-1$)
and bound pairs ($S^z=-2$) as the Hilbert space.
To manage the state overlap between two magnons and a bound pair,
we extend the Hilbert space and
introduce an effective repulsive interaction in the Hamiltonian.
The creation operator for a bound pair is expressed as
$b_{\bm k}^\dagger=(N_\Lambda)^{-1/2}\sum_{\bm q} g_{\bm q} a_{{\bm k}/2+{\bm q}}^\dagger a_{{\bm k}/2-{\bm q}}^\dagger$.
Here $a_{\bm k}^\dagger$ denotes the bosonic creation operator of a magnon with momentum
${\bm k}$, $N_\Lambda$ is the number of the lattice sites, and $g_{\bm q}$ is a form factor
satisfying the normalization $2 (N_\Lambda)^{-1}\sum_{\bm q} |g_{\bm q}|^2=1$.
We use the form factor derived from the exact bound two-magnon eigenstate in the lowest-energy mode above
the fully polarized state,
available through various methods~\cite{Wortis1963,Mattis2006,ZhitomirskyT2010,UedaM2013}.
In spin-nematic phases in frustrated
ferromagnets on the zigzag chain~\cite{KeckeMF}, the square lattice~\cite{ShannonMS},
and the body-centered-cubic lattice \cite{UedaM2013}, the function $g_{\bm q}$ is real, whereas, in
a frustrated ferromagnet on the triangular lattice, $g_{\bm q}$ can be complex due to a chiral degeneracy~\cite{MomoiSK2012}.

We investigate an interacting boson system with these two types of magnetic excitations,
characterized by the Hamiltonian
\begin{align}\label{eq:H}
{\cal H} &=\sum_{\bm k} \varepsilon_{1,{\bm k}} a_{\bm k}^\dagger a_{\bm k}
+\frac{1}{2N_\Lambda} \sum_{{\bm k}, {\bm k}^\prime, {\bm q}} v_{1,{\bm q}}
a^\dagger_{{\bm k}+{\bm q}} a^\dagger_{{\bm k}^\prime-{\bm q}} a_{{\bm k}} a_{{\bm k}^\prime}\nonumber\\
&+ \sum_{\bm k} \varepsilon_{2,{\bm k}} b_{\bm k}^\dagger b_{\bm k}
+\frac{1}{2N_\Lambda} \sum_{{\bm k}, {\bm k}^\prime, {\bm q}} v_{2,{\bm q}}
b^\dagger_{{\bm k}+{\bm q}} b^\dagger_{{\bm k}^\prime-{\bm q}} b_{{\bm k}} b_{{\bm k}^\prime}\nonumber\\
&+\frac{1}{N_\Lambda} \sum_{{\bm k}, {\bm k}^\prime, {\bm q}} u_{\bm q}
a^\dagger_{{\bm k}+{\bm q}} a_{{\bm k}} b^\dagger_{{\bm k}^\prime-{\bm q}} b_{{\bm k}^\prime}.
\end{align}
Here, $\varepsilon_{1,{\bm k}}$ ($\varepsilon_{2,{\bm k}}$) and $v_{1,{\bm q}}$ ($v_{2,{\bm q}}$)
denote the excitation energy and the repulsive potential of one-magnons (bound bi-magnons), respectively,
while $u_{\bm q}$ represents the repulsion between magnons and bi-magnons.
These interactions encompass both microscopic interactions and the effect of removing the extended excess boson space.
A similar approach is adopted in the flavor wave expansion, where repulsive interactions are included by expanding Holstein-Primakoff type square root
operators~\cite{Papanicolaou1984,SmeraldUedaShannon2015}.
The energies $\varepsilon_{1,{\bm k}}$ and $\varepsilon_{2,{\bm k}}$ include the
Zeeman energies $h$ and $2 h$, respectively, with explicit expressions:
$\varepsilon_{1,{\bm k}}=\varepsilon_{1,{\bm k}}^{(0)}+h$ and
$\varepsilon_{2,{\bm k}}=\varepsilon_{2,{\bm k}}^{(0)}+2h$.
The Hamiltonian with only bi-magnon operators $b$ has been previously discussed in studies of spin-nematic states\cite{KeckeMF,HikiharaKMF,SmeraldUedaShannon2015},
while the Hamiltonian involving both types of bosons has been utilized to describe
the spin-nematic Tomonaga-Luttinger liquid~\cite{Ramos2018}.

Here, we briefly delve into the generality of the Hamiltonian form.
The original spin system maintains spin U(1) rotation symmetry around a magnetic field.
During rotation, the boson operators $a$ and $b$ undergo a phase change to $ae^{i\theta}$ and $be^{2i\theta}$, respectively.
Due to the U(1) symmetry, interactions involving the same kind of bosons must have an equal number of creation and
annihilation operators. Interactions between different boson species are allowed, provided they satisfy U(1) invariance.
For example, $a^\dagger a b^\dagger b$ and $a^\dagger a^\dagger b$.
In our study, we omitted pair creation and annihilation terms such as $a a b^\dagger$ and $a^\dagger a^\dagger b$
for simplicity,
as they only quantitatively alter the physical quantities in our mean-field treatment.

To simplify the analysis, we disregard the momentum dependences in the potentials, specifically setting
$v_{1,{\bm q}}=v_1$, $v_{2,{\bm q}}=v_2$, and $u_{\bm q}=u$.
Thus, microscopic interactions are approximated with local on-site effective repulsions.
Since we focus on the low-density regime, where bosonic particles
are well separated each other, microscopic interaction details are not crucial;
instead, they can be represented by on-site effective couplings.
The remaining information from the microscopic models are incorporated in the energy spectra
$\varepsilon_{1,{\bm k}}$ and $\varepsilon_{2,{\bm k}}$, as well as in the form factor
$g_{\bm q}$ of bi-magnons in this framework.
Additionally, for simplicity, we assume the space inversion symmetry, implying
$\varepsilon_{1,-{\bm k}}=\varepsilon_{1,{\bm k}}$ and
$\varepsilon_{2,-{\bm k}}=\varepsilon_{2,{\bm k}}$.

\begin{figure}[thb]
  \centering
  \includegraphics[width=7cm]{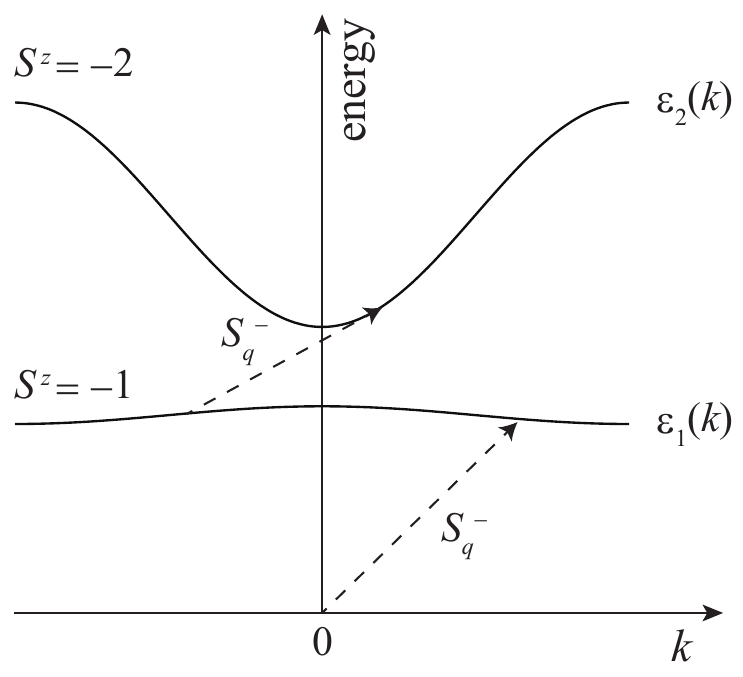}
  \caption{Energy spectrum for single-particle excitations with $S^z=-1$ and
  bound excitation pairs with $S^z=-2$ in a gapped phase. The arrows represent selected matrix elements of
  the spin lowering operator $S_q^-$.}
  \label{fig:spectrum}
\end{figure}

In the extended boson Hilbert space, we have determined the matrix elements of spin operators
in a low-density regime. For $S=1/2$~\cite{Imajo}, we find that
\begin{align}\label{eq:spin_opFMa}
  S_{\bm k}^- \simeq & a_{\bm k}^\dagger
  + \frac{2}{\sqrt{N_\Lambda}}\sum_{\bm q}
  g^\ast_{({\bm k}-{\bm q}) / 2}
  b_{{\bm k}+{\bm q}}^\dagger a_{\bm q} + \dots,\\
  S_{\bm k}^z \simeq & {\sqrt{N_\Lambda}\over 2} \delta_{{\bm k},0}
  -\frac{1}{\sqrt{N_\Lambda}} \sum_{\bm q} ( a_{{\bm k}+{\bm q}}^\dagger a_{\bm q}
  +
   2 b_{{\bm k}+{\bm q}}^\dagger b_{\bm q} ),
\label{eq:spin_opFMb}
\end{align}
while for arbitrary spin $S$,
\begin{align}\label{eq:spin_opFM2a}
  S_{\bm k}^- \simeq & \sqrt{2S} a_{\bm k}^\dagger
  + \frac{2\sqrt{2S}}{\sqrt{N_\Lambda}}\sum_{\bm q}
  g^\ast_{({\bm k}-{\bm q}) / 2}
  b_{{\bm k}+{\bm q}}^\dagger a_{\bm q} \nonumber\\
  +& \frac{2(\sqrt{2S-1}-\sqrt{2S})}{\sqrt{N_\Lambda}} g^\ast (0) \sum_{\bm q}
  b_{{\bm k}+{\bm q}}^\dagger a_{\bm q}
  + \dots,\\
  S_{\bm k}^z \simeq & S \sqrt{N_\Lambda} \delta_{{\bm k},0}
  -\frac{1}{\sqrt{N_\Lambda}} \sum_{\bm q} ( a_{{\bm k}+{\bm q}}^\dagger a_{\bm q}
  +
   2 b_{{\bm k}+{\bm q}}^\dagger b_{\bm q} ),
\label{eq:spin_opFM2b}
\end{align}
where $g (0)={N_\Lambda}^{-1} \sum_{\bm q} g_{\bm q}$ denotes
the on-site form factor of bi-magnons, and
$g(0)=0$ for $S=1/2$ due to the hard-core on-site repulsion.
In the expression for $S_{\bm k}^-$, we have omitted terms with three or more operators.
The terms with $b_{{\bm k}+{\bm q}}^\dagger a_{\bm q}$
represent transitions from a single magnon to a bound pair.
An illustration of these transitions is shown in Fig.~\ref{fig:spectrum}.
The mapping between spin and boson operators in $S_{\bm k}^z$ was used
to construct the bosonic low-energy effective theory
for the spin-nematic Tomonaga-Luttinger liquid in one dimension~\cite{KeckeMF},
where the field-theoretical prediction showed
an excellent agreement with the numerical results obtained from the corresponding $S=1/2$ spin model~\cite{HikiharaKMF}.
Additionally, for the $S\geq 1$ case, approximating the bi-magnon state as consisting solely
of two magnons on the same site (i.e., $g({\bm k})=1/\sqrt2 $)
makes the mappings (\ref{eq:spin_opFM2a}) and (\ref{eq:spin_opFM2b}) equivalent to one
form of the flavor-wave expansions~\cite{Papanicolaou1984,SmeraldUedaShannon2015}.
However, we note that, in the exact two-magnon bound states at the saturation field of quadrupolar phases,
the two magnons are usually spatially spread out and distributed at nearby sites~\cite{Penc2011,takata2015}.

We also find that a spin-pair lowering operator, important in describing the spin-nematic order parameter,
operates as
\begin{align}\label{eq:pair_op}
  & S^-_{\bm k} S^-_{{\bm k}^\prime} \simeq a^\dagger_{\bm k} a^\dagger_{{\bm k}^\prime} \nonumber\\
  & +\frac{2}{\sqrt{N_\Lambda}} \left\{2S g^\ast_{({\bm k}-{\bm k}^\prime)/2}
  +\left[\sqrt{2S(2S-1)}-2S\right] g^\ast(0) \right\} b^\dagger_{{\bm k}+{\bm k}^\prime} \nonumber\\
  & + \dots
\end{align}
in the low-density limit.
Establishing a mapping between spin and boson operators enables us to use various tools and existing results in the framework of interacting boson theory to study spin-nematic states.

\subsubsection{Bi-magnon BEC}

We consider the case in which, as the magnetic field $h$ is decreased, bound pairs first close the excitation gap
at a critical field $h_{\rm c}$ at the $\Gamma$ point ${\bm k}={\bm 0}$.
At this critical field, we have
$\varepsilon_{2,{\bm 0}}^{(0)}+2h_{\rm c}=0$,
while the single particle excitations still have a positive energy gap with
$\varepsilon_{1,{\bm k}}^{(0)}+h_{\rm c}>0$ for any ${\bm k}$.
Following the closure of the excitation energy gap, the bound pairs show the Bose-Einstein condensation (BEC).
In this regime, the bound-pair creation operator is written as
\begin{align}
  b_{\bm k}^\dagger & = \sqrt{N_{\rm c}} e^{-2i\theta} \delta_{{\bm k},0}+\tilde{b}_{\bm k}^\dagger,
\end{align}
where $N_{\rm c}$ denotes the number of condensed bound pairs.
By using this relation, the spin operators are expressed, for example, in the case of $S=1/2$, as
\begin{align}\label{eq:spin_opFM3a}
  S_{\bm k}^- & \simeq a_{\bm k}^\dagger
  + 2 \sqrt{n_{\rm c}} g^\ast_{\bm k} a_{-{\bm k}}
  + \frac{2}{\sqrt{N_\Lambda}} \sum_{\bm q} g^\ast_{({\bm k}-{\bm q} )/ 2}
  \tilde{b}_{{\bm k}+{\bm q}}^\dagger a_{\bm q}+ \dots,\\
  S_{\bm k}^z & \simeq \left(\frac{1}{2} -2n_{\rm c} \right) \sqrt{N_\Lambda} \delta_{{\bm k},0}
  -2\sqrt{n_{\rm c}} (\tilde{b}_{-k}+\tilde{b}^\dagger_k)\nonumber\\
  & -\frac{1}{\sqrt{N_\Lambda}} \sum_{\bm q} ( a_{{\bm k}+{\bm q}}^\dagger a_{\bm q}
  +
   2 \tilde{b}_{{\bm k}+{\bm q}}^\dagger \tilde{b}_{\bm q} ),
\label{eq:spin_opFM3b}
\end{align}
where $n_{\rm c}$ denotes the condensate density $n_{\rm c}=N_{\rm c}/N_\Lambda$ of bound pairs

To describe the dilute Bose gas composed of bi-magnons and magnons, as well as bi-magnon BEC,
we use the self-consistent Hartree-Fock-Popov (HFP) approximation~\cite{fetterw1971,popov1988}, as
used in Ref.~\cite{NikuniOshikawa2000}.
Within this approximation, we obtain the mean-field Hamiltonian
${\cal H}_{\rm HFP}$ in a quadratic form.
By applying the Bogoliubov transformation, we diagonalize the Hamiltonian as
\begin{align}\label{eq:H_MF2}
 {\cal H}_{\rm HFP} & = \sum_{\bm k} E_{1,{\bm k}} a^\dagger_{\bm k}a_{\bm k}
  +\sum_{\bm k} E_{2,{\bm k}} \beta^\dagger_{\bm k} \beta_{\bm k}+const.,
\end{align}
where
\begin{align}
  E_{1,{\bm k}} & = \varepsilon_{1,{\bm k}} + 2 n_1 v_1 + n_2 u, \label{eq:E1}\\
  E_{2,{\bm k}} &= \sqrt{(\varepsilon_{2,{\bm k}}+2 n_2 v_2 + n_1 u)^2-(n_{\rm c} v_2 )^2},\label{eq:E2} \\
  \beta_{\bm k} &= \tilde{b}_{\bm k} \cosh \phi_{\bm k} +\tilde{b}_{-{\bm k}}^\dagger \sinh \phi_{\bm k}
\end{align}
with $\cosh 2\phi_{\bm k}=(\varepsilon_{2,{\bm k}}+2 n_2 v_2 + n_1 u ) /E_{2,{\bm k}}$ and
$\sinh 2\phi_{\bm k}=n_{\rm c} v_2 /E_{2,{\bm k}}$.
Here $n_1$ and $n_2$ denote the densities
given by the thermal averages
$n_1=\langle a^\dagger_i a_i \rangle$ and $n_2=\langle b^\dagger_i b_i \rangle$,
which are evaluated with the Hamiltonian ${\cal H}_{\rm HFP}$ at a finite temperature.
The bi-magnon and condensate densities, $n_2$ and $n_{\rm c}$, satisfy
$n_2-n_{\rm c}=\langle \tilde{b}_i^\dagger \tilde{b}_i\rangle$,
i.e.,
\begin{align}\label{eq:SCE1_1}
 & n_2-n_{\rm c} \nonumber\\
 & = \frac{1}{N_\Lambda}\sum_{\bm k} \left\{
 \frac{ \varepsilon_{2,{\bm k}}+2v_2 n_2 +u n_1 }{E_{2,{\bm k}}}\left[ n_{\rm B} (E_{2,{\bm k}})+\frac{1}{2} \right]
 -\frac{1}{2} \right\} .
\end{align}
The function $n_{\rm B} (E)$ denotes the Bose distribution function $[\exp(E/k_{\rm B} T)-1 ]^{-1}$.
Furthermore, the HFP approximation~\cite{popov1988} imposes the condition
\begin{align}\label{eq:SCE1_2}
  \varepsilon_{2,{\bm 0}} +(2n_2-n_{\rm c}) v_2+ n_1 u=0.
\end{align}
Consequently, by solving these relations self-consistently, we determine the values of $n_1$, $n_2$, and $n_{\rm c}$
 at a finite temperature.

The one-magnon excitation energy $E_{1,{\bm k}}$ undergoes corrections arising from
finite densities and interactions.
A similar calculation has been done
in a one-dimensional system~\cite{Ramos2018}.
In the condensed phase ($n_{\rm c}>0$), $E_{2,{\bm k}}$ becomes gapless at ${\bm k}={\bm 0}$
due to relation (\ref{eq:SCE1_2}), and
it exhibits a $k$-linear dispersion relation of the Nambu-Goldstone mode near ${\bm k}={\bm 0}$
with the velocity $v_{\rm NG}=\sqrt{v_2 n_{\rm c}/M}$, where $M$ is given
by $\varepsilon_{2,{\bm k}} \simeq (2M)^{-1} {\bm k}^2+2(h-h_{\rm c})$ in the limit of small wave vectors.
In the high-temperature non-condensed phase ($n_{\rm c}=0$), $E_{2,{\bm k}}$ exhibits a positive energy gap,
which vanishes at a transition temperature with decreasing temperature.
Hereafter, we focus on situations where the dispersion relation
$E_{1,{\bm k}}$ possesses a positive energy gap $\Delta_1$, i.e., $E_{1,{\bm k}}\ge \Delta_1>0$.

\begin{figure}
  \centering
  \includegraphics[width=7.5cm]{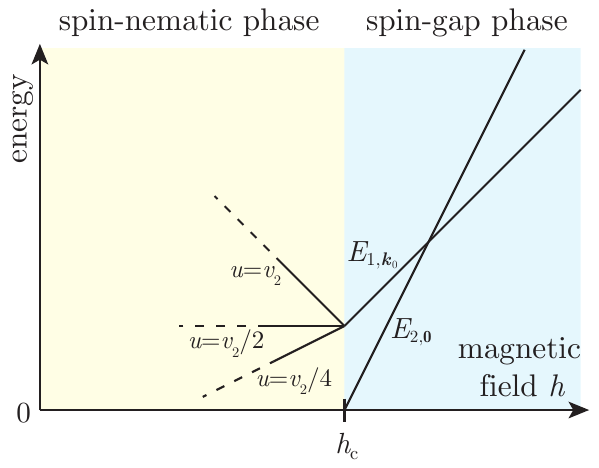}
  \caption{Field dependence of the lowest energy of one-magnon excitations ($E_{1,{\bm k}_0}$)
  and bi-magnon excitations ($E_{2,{\bm 0}}$). Inside the spin-nematic phase,
  the slope of the one-magnon gap changes depending on the interaction ratio
  $u/v_2$.}\label{fig:e_vs_h}
\end{figure}

The field dependence of the excitation gaps at $T=0$ is shown in Fig.~\ref{fig:e_vs_h}.
In the spin-gap ground state, the one-magnon excitation energy, Eq.~(\ref{eq:E1}), results in
$E_{1,{\bm k}}=\varepsilon_{1,{\bm k}}^{(0)}+h$ since both $n_1$ and $n_2$ are zero.
In the spin-nematic phase, the energy includes a correction from the finite density $n_2$ and interactions;
at $T=0$, the one-magnon density $n_1$ is found to vanish since one-magnons have a gap.
The densities $n_2$ and $n_c$, which usually satisfy $n_2>n_c$ due to zero-point oscillation,
can be calculated from Eqs.~(\ref{eq:SCE1_1}) and (\ref{eq:SCE1_2}) at zero temperature.
In the limit of low density $n_2\rightarrow 0+$ for bi-magnons,
as $h\rightarrow h_{\rm c}-$, the condensate density approaches the total density,
resulting in $n_{\rm c}/n_2 \rightarrow 1$.
In the case $n_{\rm c}=n_2$, one finds
$\varepsilon_{2,{\bm 0}}+n_2 v_2=0$, leading to the field dependence of
the one-magnon energy gap expressed as $ E_{1,{\bm k}_0} 
=\varepsilon_{1,{\bm k}_0}^{(0)}+h_{\rm c}+(1-2u/v_2) (h-h_{\rm c})$.
Here, $\varepsilon_{1,{\bm k}_0}^{(0)}+h_{\rm c}$ is the one-magnon gap at the critical field $h_{\rm c}$,
and below $h_{\rm c}$, the slope as a function of $h$ changes to $1-2u/v_2$.
Hence, the interaction effect
in the spin-nematic phase impedes the rapid closure of the one-magnon gap and may even
widen the gap.

In the case of $u<v_2/2$, the one-magnon gap closes as the magnetic field decreases in the spin-nematic phase.
This closure leads to an additional phase transition, resulting in magnon BEC and
a transition to an antiferromagnetic phase.
This phenomenon has been observed in a two-dimensional hard-core boson system with correlated
hopping~\cite{Bendjama2005,Schmidt2006} and a two-dimensional ferromagnetic $J_1$-antiferromagnetic $J_2$
model~\cite{Jiang2023},
indicating that these systems are effectively described by the model with $u<v_2/2$.
Conversely, in the case of $u>v_2/2$, the one-magnon gap begins to increase at the critical field.
Numerical calculations of the zigzag chain showed a weak gap increase below
saturation with decreasing field~\cite{SatoHM2013}, implying $u>v_2/2$.
Experimental observations on triplon excitations in SrCu$_2$(BO$_3$)$_2$,
a potential material for spin-nematic order, showed an upward turn with a kink near the magnetic
field where the magnetization begins to increase~\cite{nojiri2003,fogh2023}.
This suggests an effective description by
the model with $u>v_2/2$. (The application of our theory to the spin-nematic phase at low magnetic fields
in spin-dimer systems is discussed in Sec.~\ref{sec:dimer}.)
These results from theoretical model calculations may vary in the presence of interlayer or interchain
coupling.

\subsection{Static quantities}

The above two-component boson theory describes a spin-nematic state in a magnetic field.
The spin expectation values are expressed as
\begin{align}\label{eq:spin_exp}
 \langle S_j^x \rangle & = \langle S_j^y \rangle = 0, \nonumber\\
 \langle S_j^z \rangle & = S- n_1- 2n_2
\end{align}
at a finite temperature.
Notably, the transverse spin components (in the $x$ and $y$ directions) show no dipolar spin order at any temperature.
This result does not depend on the form factor $g_{\bm q}$.

In contrast, in the low-temperature condensed phase ($n_{\rm c}>0$), the following quadrupolar order
appears on bonds as
\begin{align}\label{eq:quad_exp}
 \langle S_i^x S_j^x - S_i^y S_j^y \rangle & = 4S\sqrt{n_c} g^\ast ({\bm r}_i-{\bm r}_j) \cos 2\theta, \nonumber\\
 \langle S_i^x S_j^y + S_i^y S_j^x \rangle & = -4S\sqrt{n_c} g^\ast ({\bm r}_i-{\bm r}_j) \sin 2\theta
\end{align}
for $i\ne j$,
where $g({\bm r})={N_\Lambda}^{-1}\sum_{\bm q}g_{\bm q}e^{-i{\bm q}\cdot {\bm r}}$,
and on the same sites for $S\ge 1$ as
\begin{align}\label{eq:quad2_exp}
 \langle (S_i^x)^2 - (S_i^y)^2 \rangle & = 2\sqrt{2S(2S-1)n_c} g^\ast (0) \cos 2\theta, \nonumber\\
 \langle S_i^x S_i^y + S_i^y S_i^x \rangle & = -2\sqrt{2S(2S-1)n_c} g^\ast (0) \sin 2\theta.
\end{align}
Note that $ \langle S_i^x S_j^x - S_i^y S_j^y \rangle$ and $\langle S_i^x S_j^y + S_i^y S_j^x \rangle$
are spin-nematic order parameters, capturing a breakdown of the spin U(1)
symmetry about the applied field~\cite{ShannonMS}.

\begin{figure}
  \centering
  \includegraphics[width=8cm]{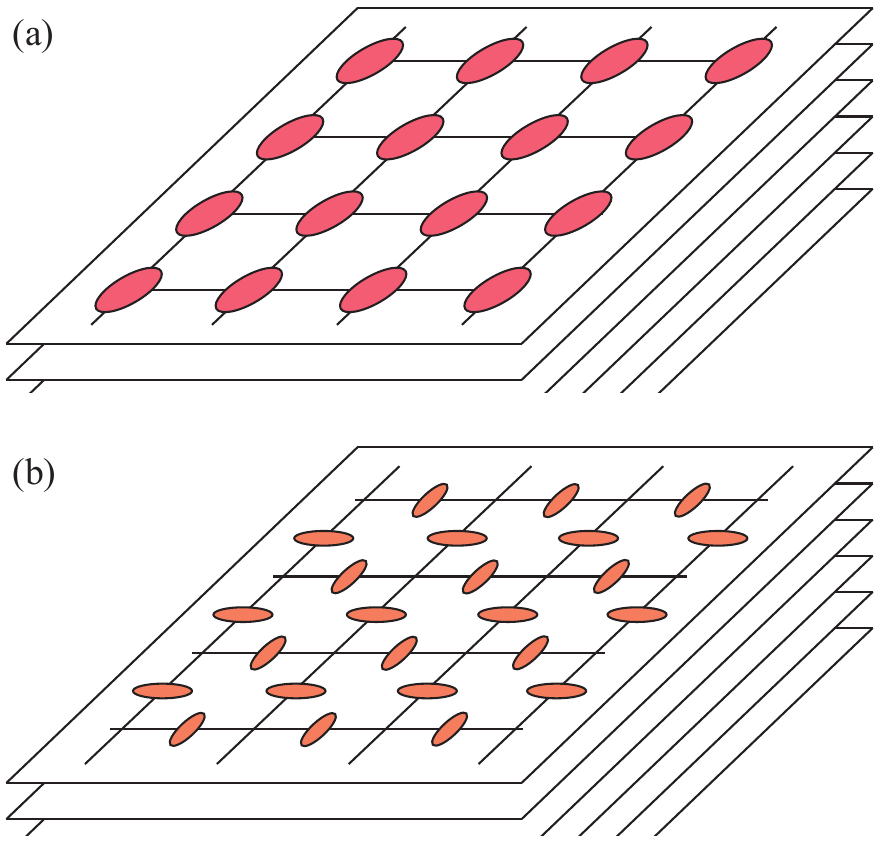}
  \caption{Illustration of quadrupolar orders on two-dimensional planes. The directors represent the arrangements of quadrupolar moments.
  (a) On-site ferro-quadrupolar moments are formed by $s$-wave magnon pairing.
  (b) Antiferro-quadrupolar moments are formed on bonds through $d$-wave paring. }\label{fig:quadrupolar}
\end{figure}

In the case the bi-magnons have $s$-wave symmetry, e.g., $g({\bm k})=1/\sqrt2$, a dominant quadrupolar order
appears on the lattice sites.
On the other hand, if the bi-magnons exhibit $d$-wave symmetry, such as
$g({\bm k})= (\cos k_x - \cos k_y )/\sqrt2$, a dominant quadrupolar order appears on the
bonds, resulting in an orthogonal director structure on two-dimensional planes.
See Fig.~\ref{fig:quadrupolar} for an illustration.

\section{Dynamical quantities in the spin-nematic state}\label{sec:dynam_SN}
Our effective theory can describe characteristic behaviors in
dynamical quantities of the spin-nematic state.
This section presents the results of the dynamical spin structure factor at zero temperature
and the NMR relaxation rate at finite temperatures.
These analyses are conducted without assuming any symmetry of the form factor of bi-magnons.

\subsection{Dynamical spin structure factor at $T=0$}\label{sec:dsf_SN}
At zero temperature, the dynamical spin structure factor is written as
\begin{align}
S^{\mu \mu} ({\bm k},\omega) &
= \pi \sum_{n} |\langle n|S_{\bm k}^\mu |0 \rangle|^2 \delta (\omega-E_n+E_0),
\end{align}
where $\mu=x,y,z$. Here, $|n\rangle$ and $E_n$ denote eigenstates and eigenenergies, with $n=0$
signifying the lowest energy level.
This quantity has been calculated for the field-induced
spin-nematic state in the spin-1 bilinear-biquadratic model in Ref.~\cite{SmeraldUedaShannon2015},
where the computation is performed under the assumption of a solely on-site $s$-wave pairing.
For comparison, we show the spin-1/2 case
with magnon pairing of arbitrary symmetry as follows:
\begin{align}\label{eq:structure}
  & S^{xx}({\bm k},\omega) = S^{yy}({\bm k},\omega) = \frac{\pi}{4} |1+2\sqrt{n_{\rm c}} g_{\bm k}|^2
  \delta (\omega-E_{1,{\bm k}})\nonumber\\
  & + \frac{\pi}{N_\Lambda} \sum_{\bm q} |g_{({\bm k}+{\bm q})/2}|^2 \sinh^2 \phi_{{\bm k}-{\bm q}}
  \delta (\omega-E_{1,{\bm q}}-E_{2,{\bm k}-{\bm q}}), \\
  & S^{zz}({\bm k},\omega) = 4 \pi n_{\rm c} e^{-2\phi_{\bm k}}
  \delta (\omega-E_{2,{\bm k}}) \nonumber\\
  & + \frac{2\pi}{N_\Lambda} \sum_{\bm q} \sinh^2 (\phi_{{\bm k}-{\bm q}}+\phi_{\bm q})
  \delta (\omega-E_{2,{\bm q}}-E_{2,{\bm k}-{\bm q}}).
\end{align}

The transverse component $S^{xx}({\bm k},\omega)$ comprises a gapful one-magnon
excitation mode with energy $E_{1,{\bm k}}$ and a continuous spectrum of two-magnon states above it.
Due to the absence of transverse spin order, this component does not exhibit
diverging singularity.
Importantly, the intensity of the one-magnon mode depends on the form factor $g_{\bm k}$ of two-magnon bound states
and the order parameter of bi-magnon BEC $|\langle b_i^\dagger\rangle|=\sqrt{n_{\rm c}}$. Thus, the
intensity
contains valuable information about the form factor.
In the case of an $s$-wave symmetric form factor, the intensity of the one-magnon mode appears
isotropic in momentum space around ${\bm k}={\bm 0}$.
Otherwise, the intensity displays oscillations in momentum space, reflecting the symmetry of
the bound magnon pairs.
Hence, this momentum-dependent signal can be a valuable tool for
detecting the spin-nematic order and identifying the symmetry
of the bound magnon pairs.

The longitudinal component $S^{zz}({\bm k},\omega)$ contains, in the lowest energy mode,
a gapless $k$-linear Nambu-Goldstone mode with energy $E_{2,{\bm k}}$, which is gapless only at ${\bm k}=0$.
Near the gapless point ${\bm k}=0$, this component behaves as $S^{zz}({\bm k},\omega)\simeq 2 \pi \sqrt{n_{\rm c} /M v_2} \norm{\bm k} \delta (\omega-v_{\rm NG} \norm{\bm k})$, and the intensity of this mode vanishes linearly.
The loss of intensity behavior at the gapless point comes from the U(1) spin symmetry of the system,
and hence, it is common in systems showing magnon BEC
(see Sec.~\ref{sec:dsf_caf}) and bi-magnon BEC~\cite{SmeraldUedaShannon2015}.
The second term in
$S^{zz}({\bm k},\omega)$ shows a continuum above the gapless mode, originating from two Nambu-Goldstone
bosons created by the operators $\beta^\dagger_{{\bm k}+{\bm q}} \beta^\dagger_{-{\bm q}}$.
We note that this continuum part does not satisfy the exact frequency sum rule $\lim_{|k|\rightarrow 0} \int_{0}^{\infty} \omega S^{zz}({\bm k},\omega) d\omega =0$,
presumably due to the Hartree-Fock approximation.

Overall, the dynamical spin structure factors do not show any diverging singularity, and the intensity
decreases linearly in the vicinity of gapless point ${\bm k}={\bm 0}$. This behavior has also been observed
in quadrupolar phases in the spin-1 bilinear-biquadratic models in
zero~\cite{Tsunetsugu2006,Penc2011,SmeraldS2013} and applied magnetic fields~\cite{SmeraldUedaShannon2015}.

\subsection{NMR relaxation rate}

\subsubsection{Formula}
Dynamical spin correlations at finite temperatures give the NMR relaxation rate $1/T_1$ as
\begin{align}\label{eq:1/T1}
   & \frac{1}{T_1}  = \frac{1}{T_{1,\perp}} + \frac{1}{T_{1,\parallel}}, \\
   & \frac{1}{T_{1,\perp}} = {1 \over \hbar N_\Lambda}
  \sum_{\bm k} \int_{-\infty}^{\infty} dt e^{i\omega_0 t}
  |A^\perp_{ {\bm k}}|^2 \sum_{\alpha=x,y}
  \langle S^{\alpha}_{\bm k} (t) S^{\alpha}_{-{\bm k}} \rangle, \nonumber\\
   & \frac{1}{T_{1,\parallel}} = {1 \over \hbar N_\Lambda}
  \sum_{\bm k} \int_{-\infty}^{\infty} dt e^{i\omega_0 t}
   |A^\parallel_{ {\bm k}}|^2
  \langle \delta S^z_{\bm k} (t) \delta S^z_{-{\bm k}} \rangle,\nonumber
\end{align}
where $A^\perp_{ {\bm k}}$ and $A^\parallel_{ {\bm k}}$ denote the form factors describing the coupling between nuclear and electronic spins, $\omega_0$ the resonance frequency of nuclear spins, and
$\delta S_{\bm k}^z=S_{\bm k}^z-\langle S_{\bm k}^z \rangle$.
We neglect the momentum dependence in the form factors $A_{\perp {\bm k}}$ and
$A_{\parallel {\bm k}}$, setting $|A_{\perp {\bm k}}|^2=c_\perp$ and
$|A_{\parallel {\bm k}}|^2=c_\parallel$, and take the limit  $\omega_0\rightarrow 0$.

\subsubsection{Transverse component}
The transverse component $1/T_{1,\perp}$ comes from the dynamics of excitations created by the operator $S_{\bm k}^-$
in Eq.~(\ref{eq:spin_opFM3a}).
Since the one-magnon excitation created by $a_{\bm k}^\dagger$ has an energy gap,
it can not receive the small energy $\hbar \omega_0$ from nuclear spins and
does not affect the relaxation.
If there is an energy overlap between one-magnon excitations and gapless collective modes, their scattering process, such as those involving integral
$\int_{-\infty}^{\infty}dt \langle \beta_{k+q}^\dagger (t) a_q(t)a_q^\dagger \beta_{k+q}\rangle$,
contributes to the relaxation rate~\cite{Imajo}.
Because of the one-magnon excitation gap $\Delta_1$, this process shows an exponential decay form $\exp(-\Delta_1 /k_{\rm B} T)$ for low temperatures
$k_{\rm B}T\ll \Delta_1$.
Three magnon terms with the form $a^\dagger a^\dagger a$ or $a^\dagger a a$
in $S^x_{\bm k}$, omitted in Eqs.~(\ref{eq:spin_opFMa}) and (\ref{eq:spin_opFM2a}), also
result in the exponential form in the relaxation rate.

\subsubsection{Longitudinal component}

The longitudinal component $1/T_{1,\parallel}$ originates from excitations created by the operator
$S_{\bm k}^z$ in Eq.~(\ref{eq:spin_opFM2b}).
This operator includes a linear term of bi-magnon operators,
contributing to a one-particle propagator in the formula for $1/T_{1,\parallel}$.
However, since the linear term in terms of the operators $\beta$ and $\beta^\dagger$ has a very weak coefficient
near ${\bm k}={\bm 0}$ as
$S^z_{\bm k} \sim {n_{\rm c}}^{1/4}\norm{\bm k}^{1/2} (\beta^\dagger_{\bm k}+\beta_{-{\bm k}})$,
the contribution from the one-particle propagator in $1/T_{1,\parallel}$ vanishes
in the limit of $\omega_0\rightarrow 0$.

Among the various terms in the longitudinal component, the dominant contribution arises from the Raman process
induced by the operator $\beta_{{\bm k}+{\bm q}}^\dagger \beta_{\bm q}$ in $S_{\bm k}^z$,
similar to the case of antiferromagnets~\cite{Moriya1956A}.
This process leads to a relaxation rate given by
\begin{align}\label{eq:1overT1_SN}
  {1 \over T_{1,\parallel}} = \frac{8\pi c_\parallel}{\hbar} \int dE & [ D_2 (E) ]^2
  \left[ 1+\frac{( n_{\rm c} v_2)^2}{E^2} \right] \nonumber\\
 &\times [ 1+n_{\rm B} (E) ] n_{\rm B} (E),
\end{align}
where $D_2(E)$ represents the density of states per site for bosons created by $\beta^\dagger$.
We note that this relaxation rate equation has the same form as Eq.~(\ref{eq:CAF_z}) derived from
the longitudinal component for
the canted antiferromagnetic phase.
In the long-wavelength approximation, the leading term results in a $T^3$ temperature dependence,
\begin{align}
  {1 \over T_{1,\parallel}} & \simeq \frac{2 c_\parallel}{3\pi \hbar v n_{\rm c}}
  \left( M k_{\rm B} T\right)^3,
\end{align}
in the low-temperature range $\hbar \omega_0 \ll k_{\rm B}T \ll k_{\rm B} T_{\rm SN}$, reflecting the $k$-linear behavior of the collective mode.
Here 
$T_{\rm SN}$ denotes the spin-nematic phase transition temperature.
As a result, the NMR relaxation rate $1/T_1$ in the spin-nematic phase exhibits a $T^3$ dependence
at low temperatures.
Notably, this finding differs from the previous studies~\cite{ShindouYM2013,SmeraldS2016}.
This most dominant behavior arises from a term that was previously omitted.

\begin{figure}
  \centering
  \includegraphics[width=8cm]{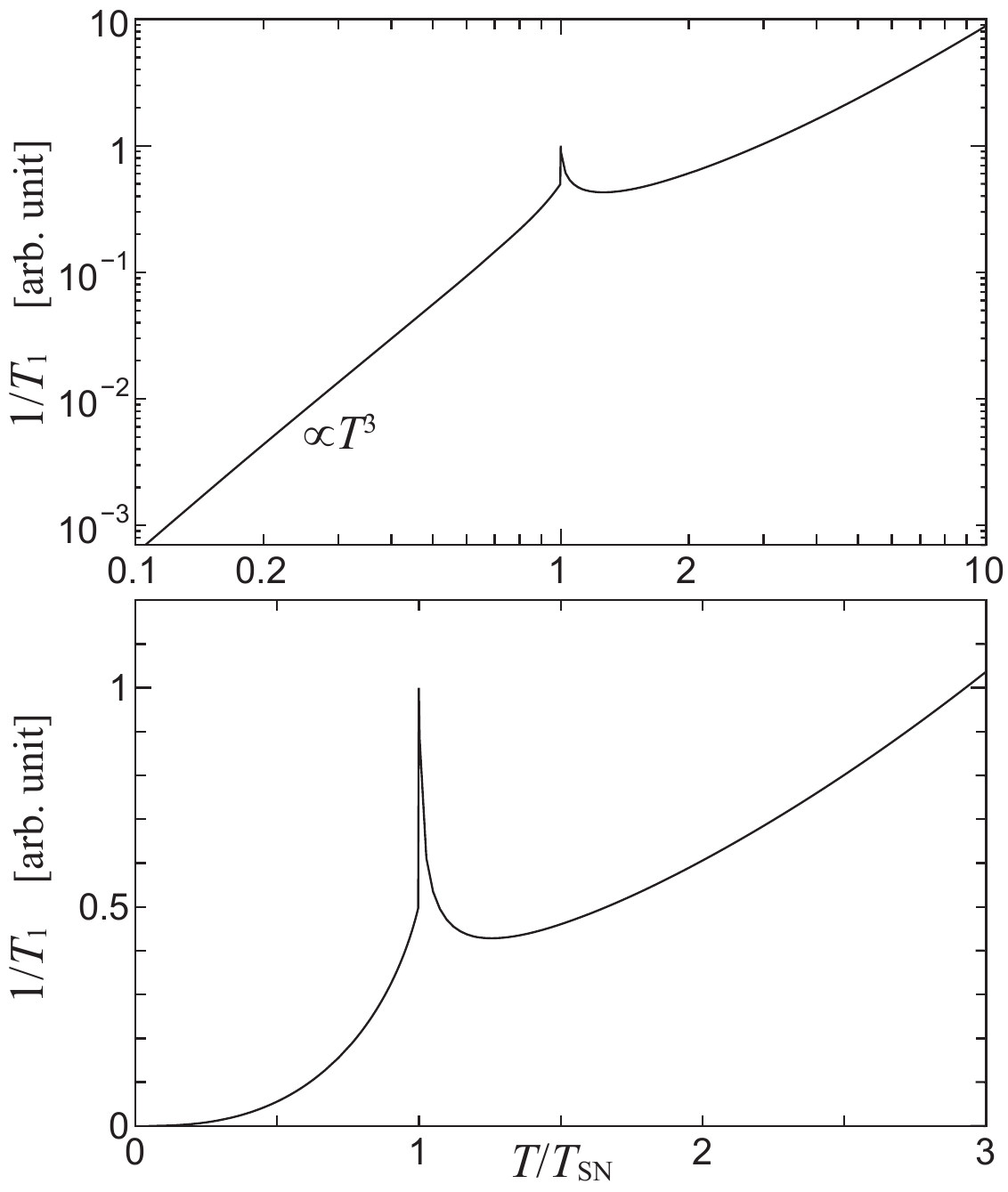}
  \caption{Temperature dependence of the NMR relaxation rate $1/T_1$ near the transition temperature
  $T_{\rm SN}$ of a spin-nematic phase in (a) logarithmic scales and (b) linear scales. The plot is
  evaluated from Eq.~(\ref{eq:1overT1_SN}). The system continuously changes to the critical point ($T=T_{\rm SN}$)
  with decreasing
  the temperature and, just below the critical temperature, the system shows a first-order phase transition
  to a spin-nematic ordered state.
  The value does not diverge at $T=T_{\rm SN}$.}\label{fig:NMR}
\end{figure}

Based on the preceding discussion, the dominant contribution to the NMR relaxation rate in
the spin-nematic phase comes from the longitudinal component presented in Eq.~(\ref{eq:1overT1_SN}).
To further investigate the temperature dependence,
we numerically computed Eq.~(\ref{eq:1overT1_SN}) on a three-dimensional lattice,
solving self-consistent equations at finite temperatures
while employing a long-wavelength approximation for the density of states.
The results of these computations are presented in Fig.~\ref{fig:NMR}.

Throughout the low-temperature spin-nematic phase, the evaluated values clearly demonstrate
a distinct $T^3$ temperature dependence.
At the transition temperature $T_{\rm SN}$, the energy gap in Eq.~(\ref{eq:E2}) closes, indicating
the achievement of the critical point as the temperature is decreased from above $T_{\rm SN}$.
However, when the temperature is increased from below $T_{\rm SN}$, a first-order phase transition occurs
at the transition point.
Importantly, the evaluated value of $1/T_1$ does not exhibit a diverging singularity at the transition temperature,
even as the temperature is decreased. Instead, it shows
a cusp-like behavior consistent with earlier field-theoretical research~\cite{SmeraldS2016}.
This absence of a diverging singularity represents a characteristic feature of the spin-nematic phase.

\section{Application to spin-dimer systems in a low magnetic field}\label{sec:dimer}
In this section, we discuss an application of our results to the spin-nematic phase in spin-dimer systems,
especially in the low magnetic field regime adjacent to the zero-field spin-gap phase.
For this investigation, we use the two-dimensional orthogonal dimer model, known as the Shastry-Sutherland model.
In the absence of a magnetic field, the ground state is the spin-singlet dimer state~\cite{shastry1981}.
The excitations in this state comprise triplons, characterized by low mobility~\cite{Miyahara1999},
and bound triplon pairs, which show high mobility~\cite{Knetter2000,Totsuka2001}.
Upon applying a magnetic field, triplons with $S^z=1$ and bound triplon pairs with $S^z=2$
become prominent low-energy degrees of freedom, and the bound pairs close the energy gap~\cite{MomoiT,WangB} at
a magnetic field.
Hereafter, we focus on these two types of degrees of freedom.

Due to the inherent structure of the Shastry-Sutherland lattice, a refinement in the definition of operators $a$ and $b$ is necessary.
This lattice contains two types of orthogonal dimer bonds, and each unit cell hosts two distinct dimer bonds.
The labeling of spin operators involves the unit cell position index $j$,
the dimer-type index $\eta=A, B$ in each unit cell, and the site index $m=1,2$ in each dimer, denoted
as $S_{j,\eta,m}^\alpha$ ($\alpha=x,y,z$).
A single triplon occupies a dimer bond, and
the creation operator for triplons with $S^z=1$ on an $\eta$-type dimer is written as $a_{{\bm k},\eta}^\dagger$.
Despite the existence of two branches of triplon energy modes, we focus on the lowest energy mode,
represented by a creation operator in the form $a_{\bm k}^\dagger=( a_{{\bm k},A}^\dagger+\sigma a_{{\bm k},B}^\dagger)/\sqrt{2}$
with $\sigma=1$ or $-1$ depending on the model parameter. Here we assume the lowest mode is non-degenerate.
We omit the higher energy mode
created by $ (a_{{\bm k},A}^\dagger-\sigma a_{{\bm k},B}^\dagger)/\sqrt{2}$.
Additionally, we consider the lowest-energy model of the bound triplon pairs with $S^z=2$, and
its creation operator is represented as
$b_{\bm k}^\dagger=(N_\Lambda)^{-1/2} \sum_{{\bm q},\eta,\eta^\prime} g_{{\bm q},\eta,\eta^\prime} a_{{\bm k}/2+{\bm q},\eta}^\dagger a_{{\bm k}/2-{\bm q},\eta^\prime}^\dagger$ with a form factor $g_{{\bm q},\eta,\eta^\prime}$.

In the extended boson Hilbert space, spanned by $a_{\bm k}^\dagger$ and
$b_{\bm k}^\dagger$, the low-energy effective Hamiltonian retains the same form as in Eq.~(\ref{eq:H}),
but with the opposite
magnetic field dependence: $\varepsilon_{1,{\bm k}}=\varepsilon_{1,{\bm k}}^{(0)}-h$ and
$\varepsilon_{2,{\bm k}}=\varepsilon_{2,{\bm k}}^{(0)}-2h$.
Consequently, the observed results of the excitation spectrum, as depicted in Fig.~\ref{fig:e_vs_h}, can be applied to
the Shastry-Sutherland model by reversing the sign of the magnetic field dependence.
In SrCu$_2$(BO$_3$)$_2$, a material with a Shastry-Sutherland lattice structure sharing many magnetic properties
with the Shastry-Sutherland model, the one-triplon excitation energy exhibits an upward trend with a kink
near the magnetic field where the magnetization starts to increase~\cite{nojiri2003,fogh2023}.
This suggests that the system enters into a spin-nematic phase at this magnetic field, and
the effective couplings for SrCu$_2$(BO$_3$)$_2$ satisfy $v_2/2 < u < v_2$.

To evaluate dynamical spin quantities, a refinement in the mapping between spin and boson operators is also necessary.
In the dilute regime, the matrix elements of spin operators can be expressed as follows:
\begin{align}\label{eq:spin_opDimer}
  & \frac{1}{\sqrt{2}} \left\{\sum_{m=1,2}(-1)^m S_{{\bm k},A,m}^+ +\sigma \sum_{m=1,2}(-1)^m S_{{\bm k},B,m}^+ \right\}
  \simeq a_{\bm k}^\dagger \nonumber\\
  & + \frac{2}{\sqrt{N_\Lambda}} \sum_{{\bm q}}
  \left\{\sum_{\eta=A,B} g^\ast_{({\bm k}-{\bm q})/2,\eta,\eta}
  + 2 \sigma g^\ast_{({\bm k}-{\bm q})/2,A,B } \right\} \nonumber\\
  & \hspace{1.5cm} \times b_{{\bm k}+{\bm q}}^\dagger a_{\bm q} + \dots, \\
  & \frac{1}{\sqrt{2}} \left\{\sum_{m=1,2}(-1)^m S_{{\bm k},A,m}^+ - \sigma \sum_{m=1,2}(-1)^m S_{{\bm k},B,m}^+ \right\}
  \simeq 0, \\
  & \sum_{\eta,m} S_{{\bm k},\eta,m}^z \simeq \frac{1}{\sqrt{N_\Lambda}}
  \sum_{{\bm q}} ( a_{{\bm k}+{\bm q}}^\dagger a_{\bm q}
 + 
 2 b_{{\bm k}+{\bm q}}^\dagger b_{\bm q} )
\end{align}
in the boson Hilbert space. Here, $\frac{1}{\sqrt{2}}\sum_{m}(-1)^m S_{j,\eta,m}^+$ represents the bond operator
creating a triplon with $S^z=1$ on the $\eta$-type dimer in the $j$-th unit cell.
These expressions enable the extraction of
information about dynamical spin quantities in the Shastry-Sutherland model using interacting boson theory.

\section{Comparisons with canted antiferromagnets}\label{sec:BEC_magnon}
To compare with the dynamical quantities in the spin-nematic phases,
we investigate the dynamical structure factors and the NMR relaxation rate
in canted antiferromagnets in a magnetic field within the present theoretical framework.

\subsection{Interacting boson theory of magnons}\label{sec:boson_theory}
We consider a spin-$S$ antiferromagnetic system near saturation. The magnons, which are bosonic particles
with quantum number $S^z=-1$, arise as excitations above the fully polarized state.
To study this system, we employ the interacting boson theory to describe a magnon Bose gas and magnon
BEC~\cite{MatsubaraMatsuda1956,NikuniOshikawa2000,GiamarchiT}.
We follow the approach used in Ref.~\cite{NikuniOshikawa2000}.
This method succeeded in explaining various thermodynamic properties, including
magnetization~\cite{NikuniOshikawa2000}, specific heat~\cite{MisguichO2004}, and thermal Hall conductivity~\cite{FurukawaM2020}.

The Hamiltonian for interacting bosons can be expressed as
\begin{align}\label{eq:H_boson}
{\cal H}_{\rm caf} &=\sum_{\bm k} \varepsilon_{\bm k} a_{\bm k}^\dagger a_{\bm k}
+\frac{1}{2N_\Lambda} \sum_{{\bm k}, {\bm k}^\prime, {\bm q}} v_{\bm q}
a^\dagger_{{\bm k}+{\bm q}} a^\dagger_{{\bm k}^\prime-{\bm q}} a_{{\bm k}} a_{{\bm k}^\prime},
\end{align}
where
$\varepsilon_{\bm k}$ denotes the excitation energy, including the Zeeman energy
$\varepsilon_{\bm k}=\varepsilon_{\bm k}^{(0)}+h$, and
$a_{\bm k} ^\dagger$ $(a_{\bm k})$ denotes the bosonic creation (annihilation)
operator for a magnon with momentum ${\bm k}$.
The interaction $v_{\bm q}$ denotes the repulsive potential between two magnons.
We omit the momentum dependence of repulsive coupling, setting $v_{\bm q}=v$.

The spin operators are written in terms of the boson operators
as
\begin{align}\label{eq:spin_opFM0}
  & S_{\bm k}^- \simeq \sqrt{2S} a_{\bm k}^\dagger
  - \frac{1}{2\sqrt{2S}N_\Lambda} \sum_{{\bm q},{\bm q}^{\prime}}
  a_{{\bm k}+{\bm q}}^\dagger a_{{\bm q}^{\prime}}^\dagger a_{{\bm q}+{\bm q}^{\prime}} ,\\
  & S_{\bm k}^z \simeq S\sqrt{N_\Lambda} \delta_{{\bm k},0}
  - \frac{1}{\sqrt{N_\Lambda}} \sum_{\bm q} a_{{\bm k}+{\bm q}}^\dagger a_{\bm q}
\end{align}
in the dilute limit.
In Eq.\ (\ref{eq:spin_opFM0}), we have expanded the Holstein-Primakoff
transformation~\cite{HolsteinPrimakoff}.


Let us assume that the energy spectrum $\varepsilon_{\bm k}$ has the lowest energy at a specific momentum
${\bm k}_0$.
We consider the case where bosons condense solely at a single momentum ${\bm k}_0$,
i.e., $2{\bm k}_0$ lies within the reciprocal lattice space.
Above the saturation field $h_{\rm c}$, the ground state has a positive energy gap
$\varepsilon_{{\bm k}_0}>0$. This gap closes at $h=h_{\rm c}$, satisfying the condition
$\varepsilon_{{\bm k}_0}^{(0)}+h_{\rm c}=0$.
In the condensed phase ($h<h_{\rm c}$), the interactions lead to BEC of magnons,
and the boson operators can be written as
\begin{align}
a_{\bm k}=\sqrt{N_{\rm c}}e^{i\theta} \delta_{{\bm k},{\bm k}_0}+\tilde{a}_{\bm k},
\end{align}
where $N_{\rm c}$ is the number of condensed bosons.

By treating interactions with the HF approximation and applying the Bogoliubov transformation,
we obtain the effective Hamiltonian
\begin{align}\label{eq:eff_H}
  {\cal H}_{\text{caf-HF}} &= \sum_{\bm k} ( E_{\bm k} \alpha_{\bm k}^\dagger \alpha_{\bm k}- vn )
\end{align}
with
\begin{align}
  E_{\bm k} &= \sqrt{(\varepsilon_{\bm k}+2vn)^2-(vn_{\rm c})^2}, \\
  \alpha_{\bm k} &= \tilde{a}_{\bm k} \cosh \theta_{\bm k} +\tilde{a}_{-{\bm k}}^\dagger \sinh \theta_{\bm k},
\end{align}
where $n$ denotes the particle density per site,
$n_{\rm c}$ represents the condensate density $n_{\rm c}=N_{\rm c}/N_\Lambda$, 
and $\theta_{\bm k}$ satisfies $\cosh 2\theta_{\bm k}=(\varepsilon_{\bm k}+2vn)/E_{\bm k}$ and
$\sinh 2\theta_{\bm k}=vn_{\rm c}/E_{\bm k}$.

In the presence of finite condensates ($n_{\rm c}>0$), the particle 
and condensate densities 
satisfy the relation
\begin{align}\label{eq:SCE1}
 \varepsilon_{{\bm k}_0}+(2n-n_{\rm c}) v =0
\end{align}
in the HFP approximation~\cite{popov1988}, leading to a gapless energy spectrum with $E_{{\bm k}_0}=0$.
In the absence of a condensate ($n_c=0$), there is no similar constraint on $n$.
The values of $n$ and $n_{\rm c}$ also satisfy the
self-consistent equation
$n-n_{\rm c}=(1/N_\Lambda) \sum_{\bm k} \langle \tilde{a}_{\bm k}^\dagger \tilde{a}_{\bm k} \rangle$,
where $\langle \cdots \rangle$ denotes the thermal average with the Hamiltonian ${\cal H}_{\text{caf-HF}}$.
The particle
and condensate densities can be calculated by solving these self-consistent equations at a finite temperature,
as in Refs.~\cite{NikuniOshikawa2000,MisguichO2004}.

We set $\theta=0$ without loss of generality.
By using the operators $\tilde{a}_{\bm k}$, we can express the spin operators: For $S=1/2$, we have
\begin{align}
  S_{\bm k}^- &\simeq \sqrt{N_{\rm c}}\left(1-\frac{n_{\rm c}}{2}\right)\delta_{{\bm k},{\bm k}_0}
  +(1-n_{\rm c})\tilde{a}_{\bm k}^\dagger -\frac{n_{\rm c}}{2}\tilde{a}_{2{\bm k}_0-{\bm k}}
  \nonumber\\
  &-\frac{\sqrt{n_{\rm c}}}{2\sqrt{N_\Lambda}} \sum_{{\bm q}}
  (\tilde{a}_{{\bm k}+{\bm q}}^\dagger \tilde{a}_{{\bm k}_0-{\bm q}}^\dagger
  +2 \tilde{a}_{{\bm k}+{\bm q}}^\dagger \tilde{a}_{{\bm k}_0+{\bm q}})
  ,  \label{eq:Sx2}\\
  S_{\bm k}^z &\simeq \sqrt{N_\Lambda} \left(\frac{1}{2}-n_{\rm c}\right) \delta_{k,0}
  - \sqrt{n_{\rm c}}(\tilde{a}_{{\bm k}+{\bm k}_0}^\dagger + \tilde{a}_{-{\bm k}+{\bm k}_0})
  \nonumber\\
  & -
  \frac{1}{\sqrt{N_\Lambda}} \sum_{\bm q} \tilde{a}_{{\bm k}+{\bm q}}^\dagger \tilde{a}_{\bm q},
  \label{eq:Sz2}
\end{align}
where we have omitted three-body terms in $S^-_{\bm k}$.
Both spin operators consist of linear and quadratic terms.
The transverse spin components align in the $x$ direction, and their expectation values are given by
$\langle S_i^x\rangle=\sqrt{n_{\rm c}} (1-{n_{\rm c}\over 2})e^{-i {\bm k}_0 \cdot{\bm r}_i}$ and
$\langle S_i^y\rangle=0$, where the zero-point oscillation from the quadratic terms in Eq.~(\ref{eq:Sx2}) has
been disregarded.

\subsection{Dynamical structure factors at $T=0$}\label{sec:dsf_caf}
We obtain the dynamical structure factors in the $S=1/2$ canted antiferromagnet
at zero temperature using this framework.
Here, we focus on the one-magnon excitation mode, which behaves as
\begin{align}\label{eq:structure_CAF}
  & S^{xx}({\bm k},\omega) = \frac{\pi}{4} \left(1-{3\over 2} n_{\rm c}\right)^2 e^{-2 \theta_{\bm k}}
  \delta (\omega-E_{{\bm k}}),\\
  & S^{yy}({\bm k},\omega) = \frac{\pi}{4} \left(1-{1\over 2}n_{\rm c} \right)^2 e^{2 \theta_{\bm k}}
  \delta (\omega-E_{{\bm k}}), \\
  & S^{zz}({\bm k},\omega) = \pi n_{\rm c} e^{-2 \theta_{{\bm k}+{\bm k}_0}}
  \delta (\omega-E_{{\bm k}+{\bm k}_0}).
\end{align}

In the low-energy regime, the transverse component $S^{yy}({\bm k},\omega)$
shows a diverging singularity at $({\bm k},\omega)=({\bm k}_0,0)$ as
  $ S^{yy}({\bm k},\omega) \simeq
  {1\over 2}\pi \sqrt{mv n_{\rm c}} (1-{1\over 2}n_{\rm c} )^2 \norm{{\bm k}-{\bm k}_0}^{-1}
  \delta (\omega-v_{\rm caf}\norm{{\bm k}-{\bm k}_0})$.
Here $m$ is given from the expansion of $\varepsilon({\bm k})$ near ${\bm k}={\bm k}_0$ as
$\varepsilon ({\bm k})\simeq (2m)^{-1} \norm{{\bm k}-{\bm k}_0}^2 + h-h_{\rm c}$
and $v_{\rm caf}=\sqrt{v n_{\rm c}/m}$ represents the velocity of magnon excitations.

In contrast, the intensities of the low-energy regions in $S^{xx}({\bm k},\omega)$
near $({\bm k},\omega)=({\bm k}_0,0)$
and $S^{zz}({\bm k},\omega)$ near $({\bm k},\omega)=({\bm 0},0)$
decay linearly as
  $S^{xx}({\bm k},\omega) \simeq
  {1\over 8} \pi (mv n_{\rm c})^{-1/2} (1-{3\over 2}n_{\rm c} )^2 \norm{{\bm k}-{\bm k}_0}
  \delta (\omega-v_{\rm caf} \norm{{\bm k}-{\bm k}_0})$ and
  $S^{zz}({\bm k},\omega) \simeq {1\over 2} \pi (n_{\rm c}/mv )^{1/2} \norm{\bm k}
  \delta (\omega-v_{\rm caf} \norm{\bm k} )$, respectively.
The linear decay of intensity in the longitudinal component $S^{zz}({\bm k},\omega)$ near
the gapless point ${\bm k}={\bm 0}$ is due to the conservation of $S^z_{\bm k}$
in the limit of ${\bm k}={\bm 0}$ and hence
is a common feature observed in systems with spin U(1) symmetry.

Lastly, we note that there have been discussions regarding the decay of magnons in canted antiferromagnets~\cite{ZhitomirskyC}. This decay mechanism contributes to the broadening of the one-magnon excitation peak.
However, the treatment of this issue needs further study beyond the HF approximation
and hence lies beyond the scope of the present study.

\subsection{NMR relaxation rate}
Moriya argued that the two-magnon Raman process contributes most dominantly
to the NMR relaxation rate in zero-field antiferromagnets~\cite{Moriya1956A,Moriya1956B}.
We derive the same result for canted antiferromagnets using the interacting boson theory described above.
Our conclusion is summarized in Table~\ref{tab:table1}.

\begin{table}[b]
\caption{\label{tab:table1}%
Temperature dependence of the NMR relaxation rate in the field-induced spin-nematic (SN) phase
and the canted antiferromagnetic (CAF) phase.
The NMR relaxation rate consists of transverse and longitudinal components corresponding to the first and second terms in Eq.~(\ref{eq:1/T1}), respectively.
Here, $\Delta_1$ denotes the gap of one-magnon excitations in the spin-nematic state.
The temperature is within the range of
$\hbar\omega_0 \ll k_{\rm B} T \ll k_{\rm B} T_{\rm c}$, where $T_{\rm c}$ denotes the transition temperature.
For the CAF phase, the behavior depends on the spin anisotropy: (a) when the spin anisotropy gap $\Delta_{\rm aniso}$ exceeds
the NMR frequency $\omega_0$, $\omega_0<\Delta_{\rm aniso}$, and (b) either in the presence of perfect U(1) spin symmetry or when $\Delta_{\rm aniso}<\omega_0$.
 }
\begin{ruledtabular}
\begin{tabular}{lcc}
 & SN phase & CAF phase  \\
\hline
Transverse component & $\exp(-\Delta_1 /k_{\rm B} T)$ &
  (a) $T^3$,\ \ (b) $T$\\
Longitudinal component & $T^3$ & $T^3$ \\
\end{tabular}
\end{ruledtabular}
\end{table}

\subsubsection{Longitudinal component}
The operator $S_{\bm k}^z$ shown in Eq.~(\ref{eq:Sz2}) includes a linear term with operators $a$ and $a^\dagger$.
The contribution from this term to $1/T_{1,\parallel}$ vanishes in the limit of $\omega_0 \rightarrow 0$,
similar to the extinction of the effect of the bi-magnon propagator in the longitudinal
component in the spin-nematic phase.
Among the various operators present in $S_{\bm k}^z$, the quadratic term with the operator $\alpha_{{\bm k}+{\bm q}}^\dagger \alpha_{\bm q}$ contributes to the Raman process.
After the time integration, only the Raman process remains non-zero in the calculation of
$\frac{1}{N_\Lambda} \sum_{\bm k} \int_{-\infty}^{\infty}
 \langle \delta S^z_{\bm k} (t) \delta S^z_{-{\bm k}} \rangle dt$, leading to
\begin{align}\label{eq:CAF_z}
  1/T_{1,\parallel} = \frac{2\pi c_\parallel}{\hbar} \int & dE [ D(E) ]^2
  \left[ 1+\frac{(vn_{\rm c})^2}{E^2} \right] \nonumber\\
 & \times [ 1+n_{\rm B} (E) ] n_{\rm B} (E).
\end{align}
Here $D(E)$ represents the density of states per site.
This equation has the same form as the well-known result obtained by Moriya using the spin-wave
expansion~\cite{Moriya1956B}, except that $n_{\rm c}$ is self-consistently determined in our approach.
Hence, the longitudinal part $1/T_{1,\parallel}$ shows a $T^3$ temperature dependence at low temperatures.

\subsubsection{Transverse component}\label{sec:CAF_trans}

The transverse component $1/T_{1,\perp}$ in Eq.~(\ref{eq:1/T1}) comes from propagations of excitations
created by the operators in $S_{\bm k}^-$ [Eq.~(\ref{eq:Sx2})].
It has been argued that the one-magnon propagator gives $1/T_{1,\parallel}$ a $T$-linear temperature
dependence at low temperatures~\cite{GiamarchiT}.
This result can be seen in our calculations;
since the coupling of the magnon creation and annihilation operators in $S^y_{\bm k}$ has a diverging
singularity at ${\bm k}={\bm k}_0$ as
$S^y_{\bm k}\sim  i ({n_{\rm c}}^{1/2}/ \norm{{\bm k}-{\bm k}_0})^{1/2} (\alpha_{\bm k}^\dagger - \alpha_{-{\bm k}} ) $,
the contribution from the one-magnon propagator to $1/T_{1,\perp}$
survives and results in a $T$-linear temperature
dependence even in the limit of $\omega_0 \rightarrow 0$.
However, in real materials, this effect is typically eliminated by
a spin anisotropy gap~\cite{Moriya1956A}.

To demonstrate the spin anisotropy dependence of the $T$-linear term in $1/T_{1,\perp}$,
we maintain a small but finite value for $\omega_0$ instead of
taking the limit $\omega_0 \rightarrow 0$.
We introduce a weak uniaxial anisotropy along the spin $x$ axis, which is
added to the Hamiltonian ${\cal H}_{\rm caf}$ as the term
$\eta \sum_{\bm k} \varepsilon_{\bm k}  (a_{\bm k}^\dagger a_{-{\bm k}}^\dagger +h.c.)$,
with $\eta$ being a small positive real constant.
In the ordered phase, the energy is minimized with $\theta=0$.
In the HFP approximation, the relation between $n$ and $n_{\rm c}$ in Eq.~(\ref{eq:SCE1}) is
replaced by $(1+2\eta) \varepsilon_{{\bm k}_0}+(2n-n_{\rm c})v=0$.
Consequently, the energy spectrum of magnons
is given by $E_{\rm aniso}({\bm k})=\sqrt{(\varepsilon_{\bm k}+2vn)^2-(vn_{\rm c}+2\eta \varepsilon_{\bm k})^2}$,
and it has an energy gap $\Delta_{\rm aniso}=\sqrt{2\eta (h_{\rm c}-h) vn_{\rm c}}$,
which closes at the critical temperature due to $n_{\rm c}=0$.
Using these results, we find that the contribution from the one-magnon propagator to $1/T_{1,\perp}$
behaves as
\begin{align}\label{eq:1/T1_ani}
 \frac{\pi c_\perp  D_{\rm aniso}(\omega_0)F(\omega_0)k_{\rm B} T}{\hbar (\omega_0)^2}
\end{align}
for $\hbar\omega_0 \ll k_{\rm B} T$,
where $D_{\rm aniso}(\omega)$ denotes the density of states with the energy spectrum
$E_{\rm aniso}({\bm k})$ and
$F(\omega)=[2\eta v(n_{\rm c}-4\eta n)+\sqrt{v^2(n_c-4\eta n)^2+(1-4\eta^2)\omega^2} ] / (1-4\eta^2)$.
Though $n$ and $n_{\rm c}$ implicitly depend on the temperature,
they can be well approximated at low temperatures with the values at zero temperature.
Hence, the dominant temperature dependence in Eq.~(\ref{eq:1/T1_ani}) is $T$-linear at low temperatures.

Note that $\omega_0$ is typically much smaller than
the energy scales of spin interactions.
Only in cases where $\omega_0 > \Delta_{\rm aniso}$, does
Eq.~(\ref{eq:1/T1_ani}) exhibit a $T$-linear behavior.
In many magnetic materials, where $\omega_0 < \Delta_{\rm aniso}$,
the term in Eq.~(\ref{eq:1/T1_ani}) disappears due to $D_{\rm aniso}(\omega_0)=0$.
Consequently, no $T$-linear behavior is observed in nearly the entire temperature range of the canted antiferromagnetic phase.
In such cases, the operator $\alpha_{{\bm k}+{\bm q}}^\dagger \alpha_{{\bm k}_0+{\bm q}}$ in $S_{\bm k}^-$,
which gives the Raman process, results in the most dominant $T^3$ temperature dependence in $1/T_{1,\perp}$.

Even in the presence of anisotropy, the critical divergence of $1/T_1$ arises from the one-particle propagator.
The anisotropy gap closes near the critical region, including the quantum critical point at $T=0$
and the phase transition line at finite temperatures.
Consequently, the influence of the one-magnon propagator becomes pronounced
in the critical region, contributing to the diverging singularity at $T=T_{\rm c}$
shown in Ref.~\cite{OrignacCG}.

\section{Summary and discussions}\label{sec:summary}

We have successfully formulated an interacting boson theory that incorporates both magnon and
bi-magnon degrees of freedom, establishing a comprehensive framework for describing spin-nematic states.
Our theory accommodates a wide range of spin-nematic states, encompassing site to bond spin-nematic states.
The characteristics of the spin-nematic order structure are embedded
into the form factor of the bound bi-magnons.
By utilizing well-established methods in the field of interacting boson theory,
our versatile method enables the calculation of various physical quantities.
In this paper, we applied the self-consistent HFP approximation and successfully obtained the dynamical quantities
in the spin-nematic phases.

The dynamical quantities in the spin-nematic phases exhibit distinct behaviors compared to
canted antiferromagnets.
The dynamical spin structure factors in the spin-nematic phases do not show diverging singularities
at any combination of momentum and frequency.
This behavior, which has also been previously observed in some specific models~\cite{Tsunetsugu2006,Penc2011,SmeraldS2013,ShindouYM2013,SmeraldUedaShannon2015}, is a general
characteristic of spin-nematic phases.
Furthermore, the transverse component of the dynamical structure factor displays a gapful structure,
indicating the presence of a finite binding energy for bi-magnons.
Notably, the intensity of one-magnon excitations in this transverse component is influenced by the form factor governing the structure of bound magnon pairs.
As a result, the intensity can exhibit oscillatory patterns around ${\bm k}={\bm 0}$, providing information about the
symmetry characteristics, such as $d$-wave, of the spin-nematic order parameter.

The NMR relaxation rate $1/T_1$ shows another characteristic in the spin-nematic phases.
Specifically, the relaxation rate does not show any diverging singularity; instead it shows a cusp
at the critical temperature of these phases.
This observation aligns with a previous field-theoretical argument~\cite{SmeraldS2016}.
Moreover, through our reanalysis of the low-temperature behavior of the relaxation rate,
we revealed a previously unreported $T^3$ temperature dependence to be the most dominant one in $1/T_1$.
This behavior is similar to that observed in canted antiferromagnets with weak spin anisotropy,
which is common in typical compounds.
Notably, in systems with perfect U(1) spin symmetry, canted antiferromagnets show a $T$-linear temperature dependence
at low temperatures, while the spin-nematic phases maintain the $T^3$ behavior.

Among the obtained results concerning the dynamical quantities, only the intensity of one-magnon
excitation mode in the dynamical
structure factor explicitly depends on the form factor of the bound magnon pairs.
Conversely, the remaining results exhibit universal features
of the spin-nematic states, indicating their independence from the symmetry of the magnon pairs.
It would be worth future work to explore additional quantities beyond those analyzed in this study
that hold the potential to reveal and identify magnon-pair symmetry.

In this paper, we focused on the case where the wavenumber vector of spin-nematic order is given as
${\bm k}_{\rm SN}={\bm 0}$.
Now, we discuss the case where the wavevector is non-zero.
If $2 {\bm k}_{\rm SN}$ points on one of the vertices of the reciprocal lattice, bi-magnons only condense at one wavenumber,
making our discussion mostly applicable.
In this case, the excitation energies take the same form as in Eq.~(\ref{eq:H_MF2}), but the
HFP condition, Eq.~(\ref{eq:SCE1_2}), changes to $\varepsilon_{2,{\bm k}_{\rm SN}} +(2n_2-n_{\rm c}) v_2+ n_1 u=0$,
and the bi-magnon excitation energy $E_{2,{\bm k}}$ has a gapless point
at wavenumber ${\bm k}_{\rm SN}$.
However,
the temperature dependence of the NMR relaxation rate remains unchanged,
indicating that the results shown in Table I exhibit universal behaviors in spin-nematic ordered states.
If the wavenumber ${\bm k}_{\rm SN}$ is incommensurate, as observed in the $J_1$-$J_2$ zigzag chain~\cite{KeckeMF},
the discussion goes beyond the analysis in this paper since bi-magnons can condense at two wavenumbers $\pm {\bm k}_{\rm SN}$.

Lastly, it is crucial to consider spin anisotropy when comparing with experiments on spin-nematic states.
In systems with spin anisotropy allowed by crystal structures, such as Dzyaloshinskii–Moriya interactions,
the symmetry breaking of spin-nematic order is either discrete or absent, as discussed in
Refs.~\cite{StarykhBalents2014,ZhangBatista2017}.
This significantly influences the critical properties of the spin-nematic phase transition.
Excitations have an anisotropy gap, and the temperature dependence of thermodynamic quantities varies
at very low temperatures.
When no symmetry is broken, the system undergoes a crossover
from a paramagnetic state to a spin-nematic state, and the anisotropy gap remains open,
even near the crossover temperature.
Whether symmetry breaking occurs or not needs to be considered for each individual crystalline structure of the material.

\acknowledgments
The author acknowledges fruitful discussions with Yoshimitsu Kohama and Shusaku Imajo, which motivated this study.
He also acknowledges stimulating discussions with Ryuichi Shindou, Akira Furusaki, Toshiya Hikihara,
Masahiro Sato, Nic Shannon, and Mladen Horvati\'c.
This work was supported by KAKENHI Grant No.\ JP20K03778 from the Japan Society for the Promotion of Science 
(JSPS).

\bibliography{NMR}

\begin{thebibliography}{62}%
\makeatletter
\providecommand \@ifxundefined [1]{%
 \@ifx{#1\undefined}
}%
\providecommand \@ifnum [1]{%
 \ifnum #1\expandafter \@firstoftwo
 \else \expandafter \@secondoftwo
 \fi
}%
\providecommand \@ifx [1]{%
 \ifx #1\expandafter \@firstoftwo
 \else \expandafter \@secondoftwo
 \fi
}%
\providecommand \natexlab [1]{#1}%
\providecommand \enquote  [1]{``#1''}%
\providecommand \bibnamefont  [1]{#1}%
\providecommand \bibfnamefont [1]{#1}%
\providecommand \citenamefont [1]{#1}%
\providecommand \href@noop [0]{\@secondoftwo}%
\providecommand \href [0]{\begingroup \@sanitize@url \@href}%
\providecommand \@href[1]{\@@startlink{#1}\@@href}%
\providecommand \@@href[1]{\endgroup#1\@@endlink}%
\providecommand \@sanitize@url [0]{\catcode `\\12\catcode `\$12\catcode
  `\&12\catcode `\#12\catcode `\^12\catcode `\_12\catcode `\%12\relax}%
\providecommand \@@startlink[1]{}%
\providecommand \@@endlink[0]{}%
\providecommand \url  [0]{\begingroup\@sanitize@url \@url }%
\providecommand \@url [1]{\endgroup\@href {#1}{\urlprefix }}%
\providecommand \urlprefix  [0]{URL }%
\providecommand \Eprint [0]{\href }%
\providecommand \doibase [0]{https://doi.org/}%
\providecommand \selectlanguage [0]{\@gobble}%
\providecommand \bibinfo  [0]{\@secondoftwo}%
\providecommand \bibfield  [0]{\@secondoftwo}%
\providecommand \translation [1]{[#1]}%
\providecommand \BibitemOpen [0]{}%
\providecommand \bibitemStop [0]{}%
\providecommand \bibitemNoStop [0]{.\EOS\space}%
\providecommand \EOS [0]{\spacefactor3000\relax}%
\providecommand \BibitemShut  [1]{\csname bibitem#1\endcsname}%
\let\auto@bib@innerbib\@empty
\bibitem [{\citenamefont {Andreev}\ and\ \citenamefont
  {Grishchuk}(1984)}]{AndreevG1984}%
  \BibitemOpen
  \bibfield  {author} {\bibinfo {author} {\bibfnamefont {A.}~\bibnamefont
  {Andreev}}\ and\ \bibinfo {author} {\bibfnamefont {I.}~\bibnamefont
  {Grishchuk}},\ }\bibfield  {title} {\bibinfo {title} {Spin nematics},\
  }\href@noop {} {\bibfield  {journal} {\bibinfo  {journal} {Sov. Phys. JETP}\
  }\textbf {\bibinfo {volume} {60}},\ \bibinfo {pages} {267} (\bibinfo {year}
  {1984})}\BibitemShut {NoStop}%
\bibitem [{\citenamefont {Shannon}\ \emph {et~al.}(2006)\citenamefont
  {Shannon}, \citenamefont {Momoi},\ and\ \citenamefont
  {Sindzingre}}]{ShannonMS}%
  \BibitemOpen
  \bibfield  {author} {\bibinfo {author} {\bibfnamefont {N.}~\bibnamefont
  {Shannon}}, \bibinfo {author} {\bibfnamefont {T.}~\bibnamefont {Momoi}},\
  and\ \bibinfo {author} {\bibfnamefont {P.}~\bibnamefont {Sindzingre}},\
  }\bibfield  {title} {\bibinfo {title} {Nematic order in square lattice
  frustrated ferromagnets},\ }\href
  {https://doi.org/10.1103/PhysRevLett.96.027213} {\bibfield  {journal}
  {\bibinfo  {journal} {Phys. Rev. Lett.}\ }\textbf {\bibinfo {volume} {96}},\
  \bibinfo {pages} {027213} (\bibinfo {year} {2006})}\BibitemShut {NoStop}%
\bibitem [{\citenamefont {Matveev}(1974)}]{Matveev1973}%
  \BibitemOpen
  \bibfield  {author} {\bibinfo {author} {\bibfnamefont {V.~M.}\ \bibnamefont
  {Matveev}},\ }\bibfield  {title} {\bibinfo {title} {Quantum quadrupolar
  magnetism and phase transitions in the presence of biquadratic exchange},\
  }\href@noop {} {\bibfield  {journal} {\bibinfo  {journal} {Sov. Phys. JETP}\
  }\textbf {\bibinfo {volume} {38}},\ \bibinfo {pages} {813} (\bibinfo {year}
  {1974})}\BibitemShut {NoStop}%
\bibitem [{\citenamefont {Papanicolaou}(1984)}]{Papanicolaou1984}%
  \BibitemOpen
  \bibfield  {author} {\bibinfo {author} {\bibfnamefont {N.}~\bibnamefont
  {Papanicolaou}},\ }\bibfield  {title} {\bibinfo {title} {Pseudospin approach
  for planar ferromagnets},\ }\href
  {https://doi.org/https://doi.org/10.1016/0550-3213(84)90268-2} {\bibfield
  {journal} {\bibinfo  {journal} {Nuclear Physics B}\ }\textbf {\bibinfo
  {volume} {240}},\ \bibinfo {pages} {281} (\bibinfo {year}
  {1984})}\BibitemShut {NoStop}%
\bibitem [{\citenamefont {Tsunetsugu}\ and\ \citenamefont
  {Arikawa}(2006)}]{Tsunetsugu2006}%
  \BibitemOpen
  \bibfield  {author} {\bibinfo {author} {\bibfnamefont {H.}~\bibnamefont
  {Tsunetsugu}}\ and\ \bibinfo {author} {\bibfnamefont {M.}~\bibnamefont
  {Arikawa}},\ }\bibfield  {title} {\bibinfo {title} {Spin nematic phase in s=1
  triangular antiferromagnets},\ }\href
  {https://doi.org/10.1143/JPSJ.75.083701} {\bibfield  {journal} {\bibinfo
  {journal} {J.\ Phys.\ Soc.\ Jpn.}\ }\textbf {\bibinfo {volume} {75}},\
  \bibinfo {pages} {083701} (\bibinfo {year} {2006})}\BibitemShut {NoStop}%
\bibitem [{\citenamefont {L\"auchli}\ \emph {et~al.}(2006)\citenamefont
  {L\"auchli}, \citenamefont {Mila},\ and\ \citenamefont
  {Penc}}]{LauchliMP2006}%
  \BibitemOpen
  \bibfield  {author} {\bibinfo {author} {\bibfnamefont {A.}~\bibnamefont
  {L\"auchli}}, \bibinfo {author} {\bibfnamefont {F.}~\bibnamefont {Mila}},\
  and\ \bibinfo {author} {\bibfnamefont {K.}~\bibnamefont {Penc}},\ }\bibfield
  {title} {\bibinfo {title} {Quadrupolar phases of the $s=1$
  bilinear-biquadratic heisenberg model on the triangular lattice},\ }\href
  {https://doi.org/10.1103/PhysRevLett.97.087205} {\bibfield  {journal}
  {\bibinfo  {journal} {Phys. Rev. Lett.}\ }\textbf {\bibinfo {volume} {97}},\
  \bibinfo {pages} {087205} (\bibinfo {year} {2006})}\BibitemShut {NoStop}%
\bibitem [{\citenamefont {Vekua}\ \emph {et~al.}(2007)\citenamefont {Vekua},
  \citenamefont {Honecker}, \citenamefont {Mikeska},\ and\ \citenamefont
  {Heidrich-Meisner}}]{Vekua2007}%
  \BibitemOpen
  \bibfield  {author} {\bibinfo {author} {\bibfnamefont {T.}~\bibnamefont
  {Vekua}}, \bibinfo {author} {\bibfnamefont {A.}~\bibnamefont {Honecker}},
  \bibinfo {author} {\bibfnamefont {H.-J.}\ \bibnamefont {Mikeska}},\ and\
  \bibinfo {author} {\bibfnamefont {F.}~\bibnamefont {Heidrich-Meisner}},\
  }\bibfield  {title} {\bibinfo {title} {Correlation functions and excitation
  spectrum of the frustrated ferromagnetic spin-$\frac{1}{2}$ chain in an
  external magnetic field},\ }\href
  {https://doi.org/10.1103/PhysRevB.76.174420} {\bibfield  {journal} {\bibinfo
  {journal} {Phys. Rev. B}\ }\textbf {\bibinfo {volume} {76}},\ \bibinfo
  {pages} {174420} (\bibinfo {year} {2007})}\BibitemShut {NoStop}%
\bibitem [{\citenamefont {Hikihara}\ \emph {et~al.}(2008)\citenamefont
  {Hikihara}, \citenamefont {Kecke}, \citenamefont {Momoi},\ and\ \citenamefont
  {Furusaki}}]{HikiharaKMF}%
  \BibitemOpen
  \bibfield  {author} {\bibinfo {author} {\bibfnamefont {T.}~\bibnamefont
  {Hikihara}}, \bibinfo {author} {\bibfnamefont {L.}~\bibnamefont {Kecke}},
  \bibinfo {author} {\bibfnamefont {T.}~\bibnamefont {Momoi}},\ and\ \bibinfo
  {author} {\bibfnamefont {A.}~\bibnamefont {Furusaki}},\ }\bibfield  {title}
  {\bibinfo {title} {Vector chiral and multipolar orders in the
  spin-$\frac{1}{2}$ frustrated ferromagnetic chain in magnetic field},\ }\href
  {https://doi.org/10.1103/PhysRevB.78.144404} {\bibfield  {journal} {\bibinfo
  {journal} {Phys. Rev. B}\ }\textbf {\bibinfo {volume} {78}},\ \bibinfo
  {pages} {144404} (\bibinfo {year} {2008})}\BibitemShut {NoStop}%
\bibitem [{\citenamefont {Sudan}\ \emph {et~al.}(2009)\citenamefont {Sudan},
  \citenamefont {L\"uscher},\ and\ \citenamefont {L\"auchli}}]{Sudan2009}%
  \BibitemOpen
  \bibfield  {author} {\bibinfo {author} {\bibfnamefont {J.}~\bibnamefont
  {Sudan}}, \bibinfo {author} {\bibfnamefont {A.}~\bibnamefont {L\"uscher}},\
  and\ \bibinfo {author} {\bibfnamefont {A.~M.}\ \bibnamefont {L\"auchli}},\
  }\bibfield  {title} {\bibinfo {title} {Emergent multipolar spin correlations
  in a fluctuating spiral: The frustrated ferromagnetic spin-$\frac{1}{2}$
  heisenberg chain in a magnetic field},\ }\href
  {https://doi.org/10.1103/PhysRevB.80.140402} {\bibfield  {journal} {\bibinfo
  {journal} {Phys. Rev. B}\ }\textbf {\bibinfo {volume} {80}},\ \bibinfo
  {pages} {140402(R)} (\bibinfo {year} {2009})}\BibitemShut {NoStop}%
\bibitem [{\citenamefont {Zhitomirsky}\ and\ \citenamefont
  {Tsunetsugu}(2010)}]{ZhitomirskyT2010}%
  \BibitemOpen
  \bibfield  {author} {\bibinfo {author} {\bibfnamefont {M.~E.}\ \bibnamefont
  {Zhitomirsky}}\ and\ \bibinfo {author} {\bibfnamefont {H.}~\bibnamefont
  {Tsunetsugu}},\ }\bibfield  {title} {\bibinfo {title} {Magnon pairing in
  quantum spin nematic},\ }\href {https://doi.org/10.1209/0295-5075/92/37001}
  {\bibfield  {journal} {\bibinfo  {journal} {Europhysics Letters}\ }\textbf
  {\bibinfo {volume} {92}},\ \bibinfo {pages} {37001} (\bibinfo {year}
  {2010})}\BibitemShut {NoStop}%
\bibitem [{\citenamefont {Sato}\ \emph {et~al.}(2013)\citenamefont {Sato},
  \citenamefont {Hikihara},\ and\ \citenamefont {Momoi}}]{SatoHM2013}%
  \BibitemOpen
  \bibfield  {author} {\bibinfo {author} {\bibfnamefont {M.}~\bibnamefont
  {Sato}}, \bibinfo {author} {\bibfnamefont {T.}~\bibnamefont {Hikihara}},\
  and\ \bibinfo {author} {\bibfnamefont {T.}~\bibnamefont {Momoi}},\ }\bibfield
   {title} {\bibinfo {title} {Spin-nematic and spin-density-wave orders in
  spatially anisotropic frustrated magnets in a magnetic field},\ }\href
  {https://doi.org/10.1103/PhysRevLett.110.077206} {\bibfield  {journal}
  {\bibinfo  {journal} {Phys. Rev. Lett.}\ }\textbf {\bibinfo {volume} {110}},\
  \bibinfo {pages} {077206} (\bibinfo {year} {2013})}\BibitemShut {NoStop}%
\bibitem [{\citenamefont {Jiang}\ \emph {et~al.}(2023)\citenamefont {Jiang},
  \citenamefont {Romh\'anyi}, \citenamefont {White}, \citenamefont
  {Zhitomirsky},\ and\ \citenamefont {Chernyshev}}]{Jiang2023}%
  \BibitemOpen
  \bibfield  {author} {\bibinfo {author} {\bibfnamefont {S.}~\bibnamefont
  {Jiang}}, \bibinfo {author} {\bibfnamefont {J.}~\bibnamefont {Romh\'anyi}},
  \bibinfo {author} {\bibfnamefont {S.~R.}\ \bibnamefont {White}}, \bibinfo
  {author} {\bibfnamefont {M.~E.}\ \bibnamefont {Zhitomirsky}},\ and\ \bibinfo
  {author} {\bibfnamefont {A.~L.}\ \bibnamefont {Chernyshev}},\ }\bibfield
  {title} {\bibinfo {title} {Where is the quantum spin nematic?},\ }\href
  {https://doi.org/10.1103/PhysRevLett.130.116701} {\bibfield  {journal}
  {\bibinfo  {journal} {Phys. Rev. Lett.}\ }\textbf {\bibinfo {volume} {130}},\
  \bibinfo {pages} {116701} (\bibinfo {year} {2023})}\BibitemShut {NoStop}%
\bibitem [{\citenamefont {Momoi}\ \emph {et~al.}(2012)\citenamefont {Momoi},
  \citenamefont {Sindzingre},\ and\ \citenamefont {Kubo}}]{MomoiSK2012}%
  \BibitemOpen
  \bibfield  {author} {\bibinfo {author} {\bibfnamefont {T.}~\bibnamefont
  {Momoi}}, \bibinfo {author} {\bibfnamefont {P.}~\bibnamefont {Sindzingre}},\
  and\ \bibinfo {author} {\bibfnamefont {K.}~\bibnamefont {Kubo}},\ }\bibfield
  {title} {\bibinfo {title} {Spin nematic order in multiple-spin exchange
  models on the triangular lattice},\ }\href
  {https://doi.org/10.1103/PhysRevLett.108.057206} {\bibfield  {journal}
  {\bibinfo  {journal} {Phys. Rev. Lett.}\ }\textbf {\bibinfo {volume} {108}},\
  \bibinfo {pages} {057206} (\bibinfo {year} {2012})}\BibitemShut {NoStop}%
\bibitem [{\citenamefont {Ueda}\ and\ \citenamefont {Momoi}(2013)}]{UedaM2013}%
  \BibitemOpen
  \bibfield  {author} {\bibinfo {author} {\bibfnamefont {H.~T.}\ \bibnamefont
  {Ueda}}\ and\ \bibinfo {author} {\bibfnamefont {T.}~\bibnamefont {Momoi}},\
  }\bibfield  {title} {\bibinfo {title} {Nematic phase and phase separation
  near saturation field in frustrated ferromagnets},\ }\href
  {https://doi.org/10.1103/PhysRevB.87.144417} {\bibfield  {journal} {\bibinfo
  {journal} {Phys. Rev. B}\ }\textbf {\bibinfo {volume} {87}},\ \bibinfo
  {pages} {144417} (\bibinfo {year} {2013})}\BibitemShut {NoStop}%
\bibitem [{\citenamefont {Janson}\ \emph {et~al.}(2016)\citenamefont {Janson},
  \citenamefont {Furukawa}, \citenamefont {Momoi}, \citenamefont {Sindzingre},
  \citenamefont {Richter},\ and\ \citenamefont {Held}}]{Janson2016}%
  \BibitemOpen
  \bibfield  {author} {\bibinfo {author} {\bibfnamefont {O.}~\bibnamefont
  {Janson}}, \bibinfo {author} {\bibfnamefont {S.}~\bibnamefont {Furukawa}},
  \bibinfo {author} {\bibfnamefont {T.}~\bibnamefont {Momoi}}, \bibinfo
  {author} {\bibfnamefont {P.}~\bibnamefont {Sindzingre}}, \bibinfo {author}
  {\bibfnamefont {J.}~\bibnamefont {Richter}},\ and\ \bibinfo {author}
  {\bibfnamefont {K.}~\bibnamefont {Held}},\ }\bibfield  {title} {\bibinfo
  {title} {Magnetic behavior of volborthite
  ${\mathrm{cu}}_{3}{\mathrm{v}}_{2}{\mathrm{o}}_{7}(\mathrm{OH}{)}_{2}\ifmmode\cdot\else\textperiodcentered\fi{}2{\mathrm{h}}_{2}\mathrm{O}$
  determined by coupled trimers rather than frustrated chains},\ }\href
  {https://doi.org/10.1103/PhysRevLett.117.037206} {\bibfield  {journal}
  {\bibinfo  {journal} {Phys. Rev. Lett.}\ }\textbf {\bibinfo {volume} {117}},\
  \bibinfo {pages} {037206} (\bibinfo {year} {2016})}\BibitemShut {NoStop}%
\bibitem [{\citenamefont {Momoi}\ and\ \citenamefont {Totsuka}(2000)}]{MomoiT}%
  \BibitemOpen
  \bibfield  {author} {\bibinfo {author} {\bibfnamefont {T.}~\bibnamefont
  {Momoi}}\ and\ \bibinfo {author} {\bibfnamefont {K.}~\bibnamefont
  {Totsuka}},\ }\bibfield  {title} {\bibinfo {title} {Magnetization plateaus of
  the shastry-sutherland model for
  ${\mathrm{srcu}}_{2}({\mathrm{bo}}_{3}{)}_{2}:$ spin-density wave,
  supersolid, and bound states},\ }\href
  {https://doi.org/10.1103/PhysRevB.62.15067} {\bibfield  {journal} {\bibinfo
  {journal} {Phys. Rev. B}\ }\textbf {\bibinfo {volume} {62}},\ \bibinfo
  {pages} {15067} (\bibinfo {year} {2000})}\BibitemShut {NoStop}%
\bibitem [{\citenamefont {Wang}\ and\ \citenamefont {Batista}(2018)}]{WangB}%
  \BibitemOpen
  \bibfield  {author} {\bibinfo {author} {\bibfnamefont {Z.}~\bibnamefont
  {Wang}}\ and\ \bibinfo {author} {\bibfnamefont {C.~D.}\ \bibnamefont
  {Batista}},\ }\bibfield  {title} {\bibinfo {title} {Dynamics and
  instabilities of the shastry-sutherland model},\ }\href
  {https://doi.org/10.1103/PhysRevLett.120.247201} {\bibfield  {journal}
  {\bibinfo  {journal} {Phys. Rev. Lett.}\ }\textbf {\bibinfo {volume} {120}},\
  \bibinfo {pages} {247201} (\bibinfo {year} {2018})}\BibitemShut {NoStop}%
\bibitem [{\citenamefont {Yokoyama}\ and\ \citenamefont
  {Hotta}(2018)}]{Yokoyama2018}%
  \BibitemOpen
  \bibfield  {author} {\bibinfo {author} {\bibfnamefont {Y.}~\bibnamefont
  {Yokoyama}}\ and\ \bibinfo {author} {\bibfnamefont {C.}~\bibnamefont
  {Hotta}},\ }\bibfield  {title} {\bibinfo {title} {Spin nematics next to spin
  singlets},\ }\href {https://doi.org/10.1103/PhysRevB.97.180404} {\bibfield
  {journal} {\bibinfo  {journal} {Phys. Rev. B}\ }\textbf {\bibinfo {volume}
  {97}},\ \bibinfo {pages} {180404(R)} (\bibinfo {year} {2018})}\BibitemShut
  {NoStop}%
\bibitem [{\citenamefont {Hikihara}\ \emph {et~al.}(2019)\citenamefont
  {Hikihara}, \citenamefont {Misawa},\ and\ \citenamefont
  {Momoi}}]{HikiharaSU4}%
  \BibitemOpen
  \bibfield  {author} {\bibinfo {author} {\bibfnamefont {T.}~\bibnamefont
  {Hikihara}}, \bibinfo {author} {\bibfnamefont {T.}~\bibnamefont {Misawa}},\
  and\ \bibinfo {author} {\bibfnamefont {T.}~\bibnamefont {Momoi}},\ }\bibfield
   {title} {\bibinfo {title} {Spin nematics in frustrated spin-dimer systems
  with bilayer structure},\ }\href
  {https://doi.org/10.1103/PhysRevB.100.214414} {\bibfield  {journal} {\bibinfo
   {journal} {Phys. Rev. B}\ }\textbf {\bibinfo {volume} {100}},\ \bibinfo
  {pages} {214414} (\bibinfo {year} {2019})}\BibitemShut {NoStop}%
\bibitem [{\citenamefont {B\"uttgen}\ \emph {et~al.}(2014)\citenamefont
  {B\"uttgen}, \citenamefont {Nawa}, \citenamefont {Fujita}, \citenamefont
  {Hagiwara}, \citenamefont {Kuhns}, \citenamefont {Prokofiev}, \citenamefont
  {Reyes}, \citenamefont {Svistov}, \citenamefont {Yoshimura},\ and\
  \citenamefont {Takigawa}}]{Buttge2014}%
  \BibitemOpen
  \bibfield  {author} {\bibinfo {author} {\bibfnamefont {N.}~\bibnamefont
  {B\"uttgen}}, \bibinfo {author} {\bibfnamefont {K.}~\bibnamefont {Nawa}},
  \bibinfo {author} {\bibfnamefont {T.}~\bibnamefont {Fujita}}, \bibinfo
  {author} {\bibfnamefont {M.}~\bibnamefont {Hagiwara}}, \bibinfo {author}
  {\bibfnamefont {P.}~\bibnamefont {Kuhns}}, \bibinfo {author} {\bibfnamefont
  {A.}~\bibnamefont {Prokofiev}}, \bibinfo {author} {\bibfnamefont {A.~P.}\
  \bibnamefont {Reyes}}, \bibinfo {author} {\bibfnamefont {L.~E.}\ \bibnamefont
  {Svistov}}, \bibinfo {author} {\bibfnamefont {K.}~\bibnamefont {Yoshimura}},\
  and\ \bibinfo {author} {\bibfnamefont {M.}~\bibnamefont {Takigawa}},\
  }\bibfield  {title} {\bibinfo {title} {Search for a spin-nematic phase in the
  quasi-one-dimensional frustrated magnet ${\mathrm{licuvo}}_{4}$},\ }\href
  {https://doi.org/10.1103/PhysRevB.90.134401} {\bibfield  {journal} {\bibinfo
  {journal} {Phys. Rev. B}\ }\textbf {\bibinfo {volume} {90}},\ \bibinfo
  {pages} {134401} (\bibinfo {year} {2014})}\BibitemShut {NoStop}%
\bibitem [{\citenamefont {Nawa}\ \emph {et~al.}(2017)\citenamefont {Nawa},
  \citenamefont {Takigawa}, \citenamefont {Kr\"amer}, \citenamefont
  {Horvati\ifmmode~\acute{c}\else \'{c}\fi{}}, \citenamefont {Berthier},
  \citenamefont {Yoshida},\ and\ \citenamefont {Yoshimura}}]{Nawa2017}%
  \BibitemOpen
  \bibfield  {author} {\bibinfo {author} {\bibfnamefont {K.}~\bibnamefont
  {Nawa}}, \bibinfo {author} {\bibfnamefont {M.}~\bibnamefont {Takigawa}},
  \bibinfo {author} {\bibfnamefont {S.}~\bibnamefont {Kr\"amer}}, \bibinfo
  {author} {\bibfnamefont {M.}~\bibnamefont {Horvati\ifmmode~\acute{c}\else
  \'{c}\fi{}}}, \bibinfo {author} {\bibfnamefont {C.}~\bibnamefont {Berthier}},
  \bibinfo {author} {\bibfnamefont {M.}~\bibnamefont {Yoshida}},\ and\ \bibinfo
  {author} {\bibfnamefont {K.}~\bibnamefont {Yoshimura}},\ }\bibfield  {title}
  {\bibinfo {title} {Dynamics of bound magnon pairs in the
  quasi-one-dimensional frustrated magnet ${\mathrm{licuvo}}_{4}$},\ }\href
  {https://doi.org/10.1103/PhysRevB.96.134423} {\bibfield  {journal} {\bibinfo
  {journal} {Phys. Rev. B}\ }\textbf {\bibinfo {volume} {96}},\ \bibinfo
  {pages} {134423} (\bibinfo {year} {2017})}\BibitemShut {NoStop}%
\bibitem [{\citenamefont {Orlova}\ \emph {et~al.}(2017)\citenamefont {Orlova},
  \citenamefont {Green}, \citenamefont {Law}, \citenamefont {Gorbunov},
  \citenamefont {Chanda}, \citenamefont {Kr\"amer}, \citenamefont
  {Horvati\ifmmode~\acute{c}\else \'{c}\fi{}}, \citenamefont {Kremer},
  \citenamefont {Wosnitza},\ and\ \citenamefont {Rikken}}]{Orlova2017}%
  \BibitemOpen
  \bibfield  {author} {\bibinfo {author} {\bibfnamefont {A.}~\bibnamefont
  {Orlova}}, \bibinfo {author} {\bibfnamefont {E.~L.}\ \bibnamefont {Green}},
  \bibinfo {author} {\bibfnamefont {J.~M.}\ \bibnamefont {Law}}, \bibinfo
  {author} {\bibfnamefont {D.~I.}\ \bibnamefont {Gorbunov}}, \bibinfo {author}
  {\bibfnamefont {G.}~\bibnamefont {Chanda}}, \bibinfo {author} {\bibfnamefont
  {S.}~\bibnamefont {Kr\"amer}}, \bibinfo {author} {\bibfnamefont
  {M.}~\bibnamefont {Horvati\ifmmode~\acute{c}\else \'{c}\fi{}}}, \bibinfo
  {author} {\bibfnamefont {R.~K.}\ \bibnamefont {Kremer}}, \bibinfo {author}
  {\bibfnamefont {J.}~\bibnamefont {Wosnitza}},\ and\ \bibinfo {author}
  {\bibfnamefont {G.~L. J.~A.}\ \bibnamefont {Rikken}},\ }\bibfield  {title}
  {\bibinfo {title} {Nuclear magnetic resonance signature of the spin-nematic
  phase in ${\mathrm{licuvo}}_{4}$ at high magnetic fields},\ }\href
  {https://doi.org/10.1103/PhysRevLett.118.247201} {\bibfield  {journal}
  {\bibinfo  {journal} {Phys. Rev. Lett.}\ }\textbf {\bibinfo {volume} {118}},\
  \bibinfo {pages} {247201} (\bibinfo {year} {2017})}\BibitemShut {NoStop}%
\bibitem [{\citenamefont {Grafe}\ \emph {et~al.}(2017)\citenamefont {Grafe},
  \citenamefont {Nishimoto}, \citenamefont {Iakovleva}, \citenamefont
  {Vavilova}, \citenamefont {Spillecke}, \citenamefont {Alfonsov},
  \citenamefont {Sturza}, \citenamefont {Wurmehl}, \citenamefont {Nojiri},
  \citenamefont {Rosner}, \citenamefont {Richter}, \citenamefont {Rößler},
  \citenamefont {Drechsler}, \citenamefont {Kataev},\ and\ \citenamefont
  {Büchner}}]{Grafe2017}%
  \BibitemOpen
  \bibfield  {author} {\bibinfo {author} {\bibfnamefont {H.-J.}\ \bibnamefont
  {Grafe}}, \bibinfo {author} {\bibfnamefont {S.}~\bibnamefont {Nishimoto}},
  \bibinfo {author} {\bibfnamefont {M.}~\bibnamefont {Iakovleva}}, \bibinfo
  {author} {\bibfnamefont {E.}~\bibnamefont {Vavilova}}, \bibinfo {author}
  {\bibfnamefont {L.}~\bibnamefont {Spillecke}}, \bibinfo {author}
  {\bibfnamefont {A.}~\bibnamefont {Alfonsov}}, \bibinfo {author}
  {\bibfnamefont {M.-I.}\ \bibnamefont {Sturza}}, \bibinfo {author}
  {\bibfnamefont {S.}~\bibnamefont {Wurmehl}}, \bibinfo {author} {\bibfnamefont
  {H.}~\bibnamefont {Nojiri}}, \bibinfo {author} {\bibfnamefont
  {H.}~\bibnamefont {Rosner}}, \bibinfo {author} {\bibfnamefont
  {J.}~\bibnamefont {Richter}}, \bibinfo {author} {\bibfnamefont {U.~K.}\
  \bibnamefont {Rößler}}, \bibinfo {author} {\bibfnamefont {S.-L.}\
  \bibnamefont {Drechsler}}, \bibinfo {author} {\bibfnamefont {V.}~\bibnamefont
  {Kataev}},\ and\ \bibinfo {author} {\bibfnamefont {B.}~\bibnamefont
  {Büchner}},\ }\bibfield  {title} {\bibinfo {title} {Signatures of a magnetic
  field-induced unconventional nematic liquid in the frustrated and anisotropic
  spin-chain cuprate licusbo4},\ }\href
  {https://doi.org/10.1038/s41598-017-06525-0} {\bibfield  {journal} {\bibinfo
  {journal} {Sci. Rep.}\ }\textbf {\bibinfo {volume} {7}},\ \bibinfo {pages}
  {6720} (\bibinfo {year} {2017})}\BibitemShut {NoStop}%
\bibitem [{\citenamefont {Skoulatos}\ \emph {et~al.}(2019)\citenamefont
  {Skoulatos}, \citenamefont {Rucker}, \citenamefont {Nilsen}, \citenamefont
  {Bertin}, \citenamefont {Pomjakushina}, \citenamefont {Ollivier},
  \citenamefont {Schneidewind}, \citenamefont {Georgii}, \citenamefont
  {Zaharko}, \citenamefont {Keller}, \citenamefont {R\"uegg}, \citenamefont
  {Pfleiderer}, \citenamefont {Schmidt}, \citenamefont {Shannon}, \citenamefont
  {Kriele}, \citenamefont {Senyshyn},\ and\ \citenamefont
  {Smerald}}]{Skoulatos2019}%
  \BibitemOpen
  \bibfield  {author} {\bibinfo {author} {\bibfnamefont {M.}~\bibnamefont
  {Skoulatos}}, \bibinfo {author} {\bibfnamefont {F.}~\bibnamefont {Rucker}},
  \bibinfo {author} {\bibfnamefont {G.~J.}\ \bibnamefont {Nilsen}}, \bibinfo
  {author} {\bibfnamefont {A.}~\bibnamefont {Bertin}}, \bibinfo {author}
  {\bibfnamefont {E.}~\bibnamefont {Pomjakushina}}, \bibinfo {author}
  {\bibfnamefont {J.}~\bibnamefont {Ollivier}}, \bibinfo {author}
  {\bibfnamefont {A.}~\bibnamefont {Schneidewind}}, \bibinfo {author}
  {\bibfnamefont {R.}~\bibnamefont {Georgii}}, \bibinfo {author} {\bibfnamefont
  {O.}~\bibnamefont {Zaharko}}, \bibinfo {author} {\bibfnamefont
  {L.}~\bibnamefont {Keller}}, \bibinfo {author} {\bibfnamefont
  {C.}~\bibnamefont {R\"uegg}}, \bibinfo {author} {\bibfnamefont
  {C.}~\bibnamefont {Pfleiderer}}, \bibinfo {author} {\bibfnamefont
  {B.}~\bibnamefont {Schmidt}}, \bibinfo {author} {\bibfnamefont
  {N.}~\bibnamefont {Shannon}}, \bibinfo {author} {\bibfnamefont
  {A.}~\bibnamefont {Kriele}}, \bibinfo {author} {\bibfnamefont
  {A.}~\bibnamefont {Senyshyn}},\ and\ \bibinfo {author} {\bibfnamefont
  {A.}~\bibnamefont {Smerald}},\ }\bibfield  {title} {\bibinfo {title}
  {Putative spin-nematic phase in $\mathrm{BaCdVO}({\mathrm{po}}_{4}{)}_{2}$},\
  }\href {https://doi.org/10.1103/PhysRevB.100.014405} {\bibfield  {journal}
  {\bibinfo  {journal} {Phys. Rev. B}\ }\textbf {\bibinfo {volume} {100}},\
  \bibinfo {pages} {014405} (\bibinfo {year} {2019})}\BibitemShut {NoStop}%
\bibitem [{\citenamefont {Yoshida}\ \emph {et~al.}(2017)\citenamefont
  {Yoshida}, \citenamefont {Nawa}, \citenamefont {Ishikawa}, \citenamefont
  {Takigawa}, \citenamefont {Jeong}, \citenamefont {Kr\"amer}, \citenamefont
  {Horvati\ifmmode~\acute{c}\else \'{c}\fi{}}, \citenamefont {Berthier},
  \citenamefont {Matsui}, \citenamefont {Goto}, \citenamefont {Kimura},
  \citenamefont {Sasaki}, \citenamefont {Yamaura}, \citenamefont {Yoshida},
  \citenamefont {Okamoto},\ and\ \citenamefont {Hiroi}}]{Yoshida2017}%
  \BibitemOpen
  \bibfield  {author} {\bibinfo {author} {\bibfnamefont {M.}~\bibnamefont
  {Yoshida}}, \bibinfo {author} {\bibfnamefont {K.}~\bibnamefont {Nawa}},
  \bibinfo {author} {\bibfnamefont {H.}~\bibnamefont {Ishikawa}}, \bibinfo
  {author} {\bibfnamefont {M.}~\bibnamefont {Takigawa}}, \bibinfo {author}
  {\bibfnamefont {M.}~\bibnamefont {Jeong}}, \bibinfo {author} {\bibfnamefont
  {S.}~\bibnamefont {Kr\"amer}}, \bibinfo {author} {\bibfnamefont
  {M.}~\bibnamefont {Horvati\ifmmode~\acute{c}\else \'{c}\fi{}}}, \bibinfo
  {author} {\bibfnamefont {C.}~\bibnamefont {Berthier}}, \bibinfo {author}
  {\bibfnamefont {K.}~\bibnamefont {Matsui}}, \bibinfo {author} {\bibfnamefont
  {T.}~\bibnamefont {Goto}}, \bibinfo {author} {\bibfnamefont {S.}~\bibnamefont
  {Kimura}}, \bibinfo {author} {\bibfnamefont {T.}~\bibnamefont {Sasaki}},
  \bibinfo {author} {\bibfnamefont {J.}~\bibnamefont {Yamaura}}, \bibinfo
  {author} {\bibfnamefont {H.}~\bibnamefont {Yoshida}}, \bibinfo {author}
  {\bibfnamefont {Y.}~\bibnamefont {Okamoto}},\ and\ \bibinfo {author}
  {\bibfnamefont {Z.}~\bibnamefont {Hiroi}},\ }\bibfield  {title} {\bibinfo
  {title} {Spin dynamics in the high-field phases of volborthite},\ }\href
  {https://doi.org/10.1103/PhysRevB.96.180413} {\bibfield  {journal} {\bibinfo
  {journal} {Phys. Rev. B}\ }\textbf {\bibinfo {volume} {96}},\ \bibinfo
  {pages} {180413(R)} (\bibinfo {year} {2017})}\BibitemShut {NoStop}%
\bibitem [{\citenamefont {Kohama}\ \emph {et~al.}(2019)\citenamefont {Kohama},
  \citenamefont {Ishikawa}, \citenamefont {Matsuo}, \citenamefont {Kindo},
  \citenamefont {Shannon},\ and\ \citenamefont {Hiroi}}]{Kohama}%
  \BibitemOpen
  \bibfield  {author} {\bibinfo {author} {\bibfnamefont {Y.}~\bibnamefont
  {Kohama}}, \bibinfo {author} {\bibfnamefont {H.}~\bibnamefont {Ishikawa}},
  \bibinfo {author} {\bibfnamefont {A.}~\bibnamefont {Matsuo}}, \bibinfo
  {author} {\bibfnamefont {K.}~\bibnamefont {Kindo}}, \bibinfo {author}
  {\bibfnamefont {N.}~\bibnamefont {Shannon}},\ and\ \bibinfo {author}
  {\bibfnamefont {Z.}~\bibnamefont {Hiroi}},\ }\bibfield  {title} {\bibinfo
  {title} {Possible observation of quantum spin-nematic phase in a frustrated
  magnet},\ }\href {https://doi.org/10.1073/pnas.1821969116} {\bibfield
  {journal} {\bibinfo  {journal} {PNAS}\ }\textbf {\bibinfo {volume} {116}},\
  \bibinfo {pages} {10686} (\bibinfo {year} {2019})}\BibitemShut {NoStop}%
\bibitem [{\citenamefont {Kimura}\ \emph {et~al.}(2022)\citenamefont {Kimura},
  \citenamefont {Imajo}, \citenamefont {Gen}, \citenamefont {Momoi},
  \citenamefont {Hagiwara}, \citenamefont {Ueda},\ and\ \citenamefont
  {Kohama}}]{Kimura}%
  \BibitemOpen
  \bibfield  {author} {\bibinfo {author} {\bibfnamefont {S.}~\bibnamefont
  {Kimura}}, \bibinfo {author} {\bibfnamefont {S.}~\bibnamefont {Imajo}},
  \bibinfo {author} {\bibfnamefont {M.}~\bibnamefont {Gen}}, \bibinfo {author}
  {\bibfnamefont {T.}~\bibnamefont {Momoi}}, \bibinfo {author} {\bibfnamefont
  {M.}~\bibnamefont {Hagiwara}}, \bibinfo {author} {\bibfnamefont
  {H.}~\bibnamefont {Ueda}},\ and\ \bibinfo {author} {\bibfnamefont
  {Y.}~\bibnamefont {Kohama}},\ }\bibfield  {title} {\bibinfo {title} {Quantum
  phase of the chromium spinel oxide ${\mathrm{hgcr}}_{2}{\mathrm{o}}_{4}$ in
  high magnetic fields},\ }\href {https://doi.org/10.1103/PhysRevB.105.L180405}
  {\bibfield  {journal} {\bibinfo  {journal} {Phys. Rev. B}\ }\textbf {\bibinfo
  {volume} {105}},\ \bibinfo {pages} {L180405} (\bibinfo {year}
  {2022})}\BibitemShut {NoStop}%
\bibitem [{\citenamefont {Imajo}\ \emph {et~al.}(2022)\citenamefont {Imajo},
  \citenamefont {Matsuyama}, \citenamefont {Nomura}, \citenamefont {Kihara},
  \citenamefont {Nakamura}, \citenamefont {Marcenat}, \citenamefont {Klein},
  \citenamefont {Seyfarth}, \citenamefont {Zhong}, \citenamefont {Kageyama},
  \citenamefont {Kindo}, \citenamefont {Momoi},\ and\ \citenamefont
  {Kohama}}]{Imajo}%
  \BibitemOpen
  \bibfield  {author} {\bibinfo {author} {\bibfnamefont {S.}~\bibnamefont
  {Imajo}}, \bibinfo {author} {\bibfnamefont {N.}~\bibnamefont {Matsuyama}},
  \bibinfo {author} {\bibfnamefont {T.}~\bibnamefont {Nomura}}, \bibinfo
  {author} {\bibfnamefont {T.}~\bibnamefont {Kihara}}, \bibinfo {author}
  {\bibfnamefont {S.}~\bibnamefont {Nakamura}}, \bibinfo {author}
  {\bibfnamefont {C.}~\bibnamefont {Marcenat}}, \bibinfo {author}
  {\bibfnamefont {T.}~\bibnamefont {Klein}}, \bibinfo {author} {\bibfnamefont
  {G.}~\bibnamefont {Seyfarth}}, \bibinfo {author} {\bibfnamefont
  {C.}~\bibnamefont {Zhong}}, \bibinfo {author} {\bibfnamefont
  {H.}~\bibnamefont {Kageyama}}, \bibinfo {author} {\bibfnamefont
  {K.}~\bibnamefont {Kindo}}, \bibinfo {author} {\bibfnamefont
  {T.}~\bibnamefont {Momoi}},\ and\ \bibinfo {author} {\bibfnamefont
  {Y.}~\bibnamefont {Kohama}},\ }\bibfield  {title} {\bibinfo {title}
  {Magnetically hidden state on the ground floor of the magnetic devil's
  staircase},\ }\href {https://doi.org/10.1103/PhysRevLett.129.147201}
  {\bibfield  {journal} {\bibinfo  {journal} {Phys. Rev. Lett.}\ }\textbf
  {\bibinfo {volume} {129}},\ \bibinfo {pages} {147201} (\bibinfo {year}
  {2022})}\BibitemShut {NoStop}%
\bibitem [{\citenamefont {Sato}\ \emph {et~al.}(2009)\citenamefont {Sato},
  \citenamefont {Momoi},\ and\ \citenamefont {Furusaki}}]{SatoMF}%
  \BibitemOpen
  \bibfield  {author} {\bibinfo {author} {\bibfnamefont {M.}~\bibnamefont
  {Sato}}, \bibinfo {author} {\bibfnamefont {T.}~\bibnamefont {Momoi}},\ and\
  \bibinfo {author} {\bibfnamefont {A.}~\bibnamefont {Furusaki}},\ }\bibfield
  {title} {\bibinfo {title} {Nmr relaxation rate and dynamical structure
  factors in nematic and multipolar liquids of frustrated spin chains under
  magnetic fields},\ }\href {https://doi.org/10.1103/PhysRevB.79.060406}
  {\bibfield  {journal} {\bibinfo  {journal} {Phys. Rev. B}\ }\textbf {\bibinfo
  {volume} {79}},\ \bibinfo {pages} {060406(R)} (\bibinfo {year}
  {2009})}\BibitemShut {NoStop}%
\bibitem [{\citenamefont {Shindou}\ \emph {et~al.}(2013)\citenamefont
  {Shindou}, \citenamefont {Yunoki},\ and\ \citenamefont
  {Momoi}}]{ShindouYM2013}%
  \BibitemOpen
  \bibfield  {author} {\bibinfo {author} {\bibfnamefont {R.}~\bibnamefont
  {Shindou}}, \bibinfo {author} {\bibfnamefont {S.}~\bibnamefont {Yunoki}},\
  and\ \bibinfo {author} {\bibfnamefont {T.}~\bibnamefont {Momoi}},\ }\bibfield
   {title} {\bibinfo {title} {Dynamical spin structure factors of quantum spin
  nematic states},\ }\href {https://doi.org/10.1103/PhysRevB.87.054429}
  {\bibfield  {journal} {\bibinfo  {journal} {Phys. Rev. B}\ }\textbf {\bibinfo
  {volume} {87}},\ \bibinfo {pages} {054429} (\bibinfo {year}
  {2013})}\BibitemShut {NoStop}%
\bibitem [{\citenamefont {Smerald}\ and\ \citenamefont
  {Shannon}(2016)}]{SmeraldS2016}%
  \BibitemOpen
  \bibfield  {author} {\bibinfo {author} {\bibfnamefont {A.}~\bibnamefont
  {Smerald}}\ and\ \bibinfo {author} {\bibfnamefont {N.}~\bibnamefont
  {Shannon}},\ }\bibfield  {title} {\bibinfo {title} {Theory of nmr $1/{T}_{1}$
  relaxation in a quantum spin nematic in an applied magnetic field},\ }\href
  {https://doi.org/10.1103/PhysRevB.93.184419} {\bibfield  {journal} {\bibinfo
  {journal} {Phys. Rev. B}\ }\textbf {\bibinfo {volume} {93}},\ \bibinfo
  {pages} {184419} (\bibinfo {year} {2016})}\BibitemShut {NoStop}%
\bibitem [{\citenamefont {Moriya}(1956{\natexlab{a}})}]{Moriya1956A}%
  \BibitemOpen
  \bibfield  {author} {\bibinfo {author} {\bibfnamefont {T.}~\bibnamefont
  {Moriya}},\ }\bibfield  {title} {\bibinfo {title} {{Nuclear Magnetic
  Relaxation in Antiferromagnetics}},\ }\href
  {https://doi.org/10.1143/PTP.16.23} {\bibfield  {journal} {\bibinfo
  {journal} {Prog.\ Theor.\ Phys.}\ }\textbf {\bibinfo {volume} {16}},\
  \bibinfo {pages} {23} (\bibinfo {year} {1956}{\natexlab{a}})}\BibitemShut
  {NoStop}%
\bibitem [{\citenamefont {Moriya}(1956{\natexlab{b}})}]{Moriya1956B}%
  \BibitemOpen
  \bibfield  {author} {\bibinfo {author} {\bibfnamefont {T.}~\bibnamefont
  {Moriya}},\ }\bibfield  {title} {\bibinfo {title} {{Nuclear Magnetic
  Relaxation in Antiferromagnetics, II}},\ }\href
  {https://doi.org/10.1143/PTP.16.641} {\bibfield  {journal} {\bibinfo
  {journal} {Prog.\ Theor.\ Phys.}\ }\textbf {\bibinfo {volume} {16}},\
  \bibinfo {pages} {641} (\bibinfo {year} {1956}{\natexlab{b}})}\BibitemShut
  {NoStop}%
\bibitem [{\citenamefont {Jaccarino}(1965)}]{Jaccarino1965}%
  \BibitemOpen
  \bibfield  {author} {\bibinfo {author} {\bibfnamefont {V.}~\bibnamefont
  {Jaccarino}},\ }\bibinfo {title} {Nuclear resonance in antiferromagnetics},\
  in\ \href@noop {} {\emph {\bibinfo {booktitle} {Magnetism IIA}}},\ \bibinfo
  {editor} {edited by\ \bibinfo {editor} {\bibfnamefont {G.~T.}\ \bibnamefont
  {Rado}}\ and\ \bibinfo {editor} {\bibfnamefont {H.}~\bibnamefont {Suhl}}}\
  (\bibinfo  {publisher} {Academic},\ \bibinfo {address} {New York},\ \bibinfo
  {year} {1965})\ pp.\ \bibinfo {pages} {307--355}\BibitemShut {NoStop}%
\bibitem [{\citenamefont {Smerald}\ \emph {et~al.}(2015)\citenamefont
  {Smerald}, \citenamefont {Ueda},\ and\ \citenamefont
  {Shannon}}]{SmeraldUedaShannon2015}%
  \BibitemOpen
  \bibfield  {author} {\bibinfo {author} {\bibfnamefont {A.}~\bibnamefont
  {Smerald}}, \bibinfo {author} {\bibfnamefont {H.~T.}\ \bibnamefont {Ueda}},\
  and\ \bibinfo {author} {\bibfnamefont {N.}~\bibnamefont {Shannon}},\
  }\bibfield  {title} {\bibinfo {title} {Theory of inelastic neutron scattering
  in a field-induced spin-nematic state},\ }\href
  {https://doi.org/10.1103/PhysRevB.91.174402} {\bibfield  {journal} {\bibinfo
  {journal} {Phys. Rev. B}\ }\textbf {\bibinfo {volume} {91}},\ \bibinfo
  {pages} {174402} (\bibinfo {year} {2015})}\BibitemShut {NoStop}%
\bibitem [{\citenamefont {Fogh}\ \emph {et~al.}(2023)\citenamefont {Fogh},
  \citenamefont {Nayak}, \citenamefont {Prokhnenko}, \citenamefont
  {Bartkowiak}, \citenamefont {Munakata}, \citenamefont {Soh}, \citenamefont
  {Turrini}, \citenamefont {Zayed}, \citenamefont {Pomjakushina}, \citenamefont
  {Kageyama}, \citenamefont {Nojiri}, \citenamefont {Kakurai}, \citenamefont
  {Normand}, \citenamefont {Mila},\ and\ \citenamefont {Rønnow}}]{fogh2023}%
  \BibitemOpen
  \bibfield  {author} {\bibinfo {author} {\bibfnamefont {E.}~\bibnamefont
  {Fogh}}, \bibinfo {author} {\bibfnamefont {M.}~\bibnamefont {Nayak}},
  \bibinfo {author} {\bibfnamefont {O.}~\bibnamefont {Prokhnenko}}, \bibinfo
  {author} {\bibfnamefont {M.}~\bibnamefont {Bartkowiak}}, \bibinfo {author}
  {\bibfnamefont {K.}~\bibnamefont {Munakata}}, \bibinfo {author}
  {\bibfnamefont {J.-R.}\ \bibnamefont {Soh}}, \bibinfo {author} {\bibfnamefont
  {A.~A.}\ \bibnamefont {Turrini}}, \bibinfo {author} {\bibfnamefont {M.~E.}\
  \bibnamefont {Zayed}}, \bibinfo {author} {\bibfnamefont {E.}~\bibnamefont
  {Pomjakushina}}, \bibinfo {author} {\bibfnamefont {H.}~\bibnamefont
  {Kageyama}}, \bibinfo {author} {\bibfnamefont {H.}~\bibnamefont {Nojiri}},
  \bibinfo {author} {\bibfnamefont {K.}~\bibnamefont {Kakurai}}, \bibinfo
  {author} {\bibfnamefont {B.}~\bibnamefont {Normand}}, \bibinfo {author}
  {\bibfnamefont {F.}~\bibnamefont {Mila}},\ and\ \bibinfo {author}
  {\bibfnamefont {H.~M.}\ \bibnamefont {Rønnow}},\ }\href@noop {} {\bibinfo
  {title} {Field-induced bound-state condensation and spin-nematic phase in
  srcu$_2$(bo$_3$)$_2$ revealed by neutron scattering up to 25.9 t}} (\bibinfo
  {year} {2023}),\ \Eprint {https://arxiv.org/abs/2306.07389}
  {arXiv:2306.07389} \BibitemShut {NoStop}%
\bibitem [{\citenamefont {Wortis}(1963)}]{Wortis1963}%
  \BibitemOpen
  \bibfield  {author} {\bibinfo {author} {\bibfnamefont {M.}~\bibnamefont
  {Wortis}},\ }\bibfield  {title} {\bibinfo {title} {Bound states of two spin
  waves in the heisenberg ferromagnet},\ }\href
  {https://doi.org/10.1103/PhysRev.132.85} {\bibfield  {journal} {\bibinfo
  {journal} {Phys. Rev.}\ }\textbf {\bibinfo {volume} {132}},\ \bibinfo {pages}
  {85} (\bibinfo {year} {1963})}\BibitemShut {NoStop}%
\bibitem [{\citenamefont {Mattis}(2006)}]{Mattis2006}%
  \BibitemOpen
  \bibfield  {author} {\bibinfo {author} {\bibfnamefont {D.~C.}\ \bibnamefont
  {Mattis}},\ }\href {https://doi.org/10.1142/5372} {\emph {\bibinfo {title}
  {The Theory of Magnetism Made Simple}}}\ (\bibinfo  {publisher} {World
  Scientific},\ \bibinfo {address} {Singapore},\ \bibinfo {year}
  {2006})\BibitemShut {NoStop}%
\bibitem [{\citenamefont {Kecke}\ \emph {et~al.}(2007)\citenamefont {Kecke},
  \citenamefont {Momoi},\ and\ \citenamefont {Furusaki}}]{KeckeMF}%
  \BibitemOpen
  \bibfield  {author} {\bibinfo {author} {\bibfnamefont {L.}~\bibnamefont
  {Kecke}}, \bibinfo {author} {\bibfnamefont {T.}~\bibnamefont {Momoi}},\ and\
  \bibinfo {author} {\bibfnamefont {A.}~\bibnamefont {Furusaki}},\ }\bibfield
  {title} {\bibinfo {title} {Multimagnon bound states in the frustrated
  ferromagnetic one-dimensional chain},\ }\href
  {https://doi.org/10.1103/PhysRevB.76.060407} {\bibfield  {journal} {\bibinfo
  {journal} {Phys. Rev. B}\ }\textbf {\bibinfo {volume} {76}},\ \bibinfo
  {pages} {060407(R)} (\bibinfo {year} {2007})}\BibitemShut {NoStop}%
\bibitem [{\citenamefont {Ramos}\ \emph {et~al.}(2018)\citenamefont {Ramos},
  \citenamefont {Eli\"ens},\ and\ \citenamefont {Pereira}}]{Ramos2018}%
  \BibitemOpen
  \bibfield  {author} {\bibinfo {author} {\bibfnamefont {F.~B.}\ \bibnamefont
  {Ramos}}, \bibinfo {author} {\bibfnamefont {S.}~\bibnamefont {Eli\"ens}},\
  and\ \bibinfo {author} {\bibfnamefont {R.~G.}\ \bibnamefont {Pereira}},\
  }\bibfield  {title} {\bibinfo {title} {Dynamical structure factors in the
  nematic phase of frustrated ferromagnetic spin chains},\ }\href
  {https://doi.org/10.1103/PhysRevB.98.094431} {\bibfield  {journal} {\bibinfo
  {journal} {Phys. Rev. B}\ }\textbf {\bibinfo {volume} {98}},\ \bibinfo
  {pages} {094431} (\bibinfo {year} {2018})}\BibitemShut {NoStop}%
\bibitem [{\citenamefont {Penc}\ and\ \citenamefont
  {L{\"a}uchli}(2011)}]{Penc2011}%
  \BibitemOpen
  \bibfield  {author} {\bibinfo {author} {\bibfnamefont {K.}~\bibnamefont
  {Penc}}\ and\ \bibinfo {author} {\bibfnamefont {A.~M.}\ \bibnamefont
  {L{\"a}uchli}},\ }\bibinfo {title} {Spin nematic phases in quantum spin
  systems},\ in\ \href {https://doi.org/10.1007/978-3-642-10589-0_13} {\emph
  {\bibinfo {booktitle} {Introduction to Frustrated Magnetism: Materials,
  Experiments, Theory}}},\ \bibinfo {editor} {edited by\ \bibinfo {editor}
  {\bibfnamefont {C.}~\bibnamefont {Lacroix}}, \bibinfo {editor} {\bibfnamefont
  {P.}~\bibnamefont {Mendels}},\ and\ \bibinfo {editor} {\bibfnamefont
  {F.}~\bibnamefont {Mila}}}\ (\bibinfo  {publisher} {Springer},\ \bibinfo
  {address} {Berlin},\ \bibinfo {year} {2011})\ pp.\ \bibinfo {pages}
  {331--362}\BibitemShut {NoStop}%
\bibitem [{\citenamefont {Takata}\ \emph {et~al.}(2015)\citenamefont {Takata},
  \citenamefont {Momoi},\ and\ \citenamefont {Oshikawa}}]{takata2015}%
  \BibitemOpen
  \bibfield  {author} {\bibinfo {author} {\bibfnamefont {E.}~\bibnamefont
  {Takata}}, \bibinfo {author} {\bibfnamefont {T.}~\bibnamefont {Momoi}},\ and\
  \bibinfo {author} {\bibfnamefont {M.}~\bibnamefont {Oshikawa}},\ }\href@noop
  {} {\bibinfo {title} {Nematic ordering in pyrochlore antiferromagnets:
  high-field phase of chromium spinel oxides}} (\bibinfo {year} {2015}),\
  \Eprint {https://arxiv.org/abs/1510.02373} {arXiv:1510.02373
  [cond-mat.str-el]} \BibitemShut {NoStop}%
\bibitem [{\citenamefont {Fetter}\ and\ \citenamefont
  {Walecka}(1971)}]{fetterw1971}%
  \BibitemOpen
  \bibfield  {author} {\bibinfo {author} {\bibfnamefont {A.~L.}\ \bibnamefont
  {Fetter}}\ and\ \bibinfo {author} {\bibfnamefont {J.~D.}\ \bibnamefont
  {Walecka}},\ }\href@noop {} {\emph {\bibinfo {title} {Quantum Theory of
  Many-Particle Systems}}}\ (\bibinfo  {publisher} {McGraw-Hill},\ \bibinfo
  {address} {New York},\ \bibinfo {year} {1971})\BibitemShut {NoStop}%
\bibitem [{\citenamefont {Popov}(1988)}]{popov1988}%
  \BibitemOpen
  \bibfield  {author} {\bibinfo {author} {\bibfnamefont {V.~N.}\ \bibnamefont
  {Popov}},\ }\href {https://doi.org/10.1017/CBO9780511599910} {\emph {\bibinfo
  {title} {Functional Integrals and Collective Excitations}}},\ Cambridge
  Monographs on Mathematical Physics\ (\bibinfo  {publisher} {Cambridge
  University Press},\ \bibinfo {address} {Cambridge},\ \bibinfo {year}
  {1988})\BibitemShut {NoStop}%
\bibitem [{\citenamefont {Nikuni}\ \emph {et~al.}(2000)\citenamefont {Nikuni},
  \citenamefont {Oshikawa}, \citenamefont {Oosawa},\ and\ \citenamefont
  {Tanaka}}]{NikuniOshikawa2000}%
  \BibitemOpen
  \bibfield  {author} {\bibinfo {author} {\bibfnamefont {T.}~\bibnamefont
  {Nikuni}}, \bibinfo {author} {\bibfnamefont {M.}~\bibnamefont {Oshikawa}},
  \bibinfo {author} {\bibfnamefont {A.}~\bibnamefont {Oosawa}},\ and\ \bibinfo
  {author} {\bibfnamefont {H.}~\bibnamefont {Tanaka}},\ }\bibfield  {title}
  {\bibinfo {title} {Bose-einstein condensation of dilute magnons in
  ${\mathrm{tlcucl}}_{3}$},\ }\href
  {https://doi.org/10.1103/PhysRevLett.84.5868} {\bibfield  {journal} {\bibinfo
   {journal} {Phys. Rev. Lett.}\ }\textbf {\bibinfo {volume} {84}},\ \bibinfo
  {pages} {5868} (\bibinfo {year} {2000})}\BibitemShut {NoStop}%
\bibitem [{\citenamefont {Bendjama}\ \emph {et~al.}(2005)\citenamefont
  {Bendjama}, \citenamefont {Kumar},\ and\ \citenamefont
  {Mila}}]{Bendjama2005}%
  \BibitemOpen
  \bibfield  {author} {\bibinfo {author} {\bibfnamefont {R.}~\bibnamefont
  {Bendjama}}, \bibinfo {author} {\bibfnamefont {B.}~\bibnamefont {Kumar}},\
  and\ \bibinfo {author} {\bibfnamefont {F.}~\bibnamefont {Mila}},\ }\bibfield
  {title} {\bibinfo {title} {Absence of single-particle bose-einstein
  condensation at low densities for bosons with correlated hopping},\ }\href
  {https://doi.org/10.1103/PhysRevLett.95.110406} {\bibfield  {journal}
  {\bibinfo  {journal} {Phys. Rev. Lett.}\ }\textbf {\bibinfo {volume} {95}},\
  \bibinfo {pages} {110406} (\bibinfo {year} {2005})}\BibitemShut {NoStop}%
\bibitem [{\citenamefont {Schmidt}\ \emph {et~al.}(2006)\citenamefont
  {Schmidt}, \citenamefont {Dorier}, \citenamefont {L\"auchli},\ and\
  \citenamefont {Mila}}]{Schmidt2006}%
  \BibitemOpen
  \bibfield  {author} {\bibinfo {author} {\bibfnamefont {K.~P.}\ \bibnamefont
  {Schmidt}}, \bibinfo {author} {\bibfnamefont {J.}~\bibnamefont {Dorier}},
  \bibinfo {author} {\bibfnamefont {A.}~\bibnamefont {L\"auchli}},\ and\
  \bibinfo {author} {\bibfnamefont {F.}~\bibnamefont {Mila}},\ }\bibfield
  {title} {\bibinfo {title} {Single-particle versus pair condensation of
  hard-core bosons with correlated hopping},\ }\href
  {https://doi.org/10.1103/PhysRevB.74.174508} {\bibfield  {journal} {\bibinfo
  {journal} {Phys. Rev. B}\ }\textbf {\bibinfo {volume} {74}},\ \bibinfo
  {pages} {174508} (\bibinfo {year} {2006})}\BibitemShut {NoStop}%
\bibitem [{\citenamefont {Nojiri}\ \emph {et~al.}(2003)\citenamefont {Nojiri},
  \citenamefont {Kageyama}, \citenamefont {Ueda},\ and\ \citenamefont
  {Motokawa}}]{nojiri2003}%
  \BibitemOpen
  \bibfield  {author} {\bibinfo {author} {\bibfnamefont {H.}~\bibnamefont
  {Nojiri}}, \bibinfo {author} {\bibfnamefont {H.}~\bibnamefont {Kageyama}},
  \bibinfo {author} {\bibfnamefont {Y.}~\bibnamefont {Ueda}},\ and\ \bibinfo
  {author} {\bibfnamefont {M.}~\bibnamefont {Motokawa}},\ }\bibfield  {title}
  {\bibinfo {title} {Esr study on the excited state energy spectrum of
  srcu2(bo3)2 –a central role of multiple-triplet bound states–},\ }\href
  {https://doi.org/10.1143/JPSJ.72.3243} {\bibfield  {journal} {\bibinfo
  {journal} {J.\ Phys.\ Soc.\ Jpn.}\ }\textbf {\bibinfo {volume} {72}},\
  \bibinfo {pages} {3243} (\bibinfo {year} {2003})}\BibitemShut {NoStop}%
\bibitem [{\citenamefont {Smerald}\ and\ \citenamefont
  {Shannon}(2013)}]{SmeraldS2013}%
  \BibitemOpen
  \bibfield  {author} {\bibinfo {author} {\bibfnamefont {A.}~\bibnamefont
  {Smerald}}\ and\ \bibinfo {author} {\bibfnamefont {N.}~\bibnamefont
  {Shannon}},\ }\bibfield  {title} {\bibinfo {title} {Theory of spin
  excitations in a quantum spin-nematic state},\ }\href
  {https://doi.org/10.1103/PhysRevB.88.184430} {\bibfield  {journal} {\bibinfo
  {journal} {Phys. Rev. B}\ }\textbf {\bibinfo {volume} {88}},\ \bibinfo
  {pages} {184430} (\bibinfo {year} {2013})}\BibitemShut {NoStop}%
\bibitem [{\citenamefont {{Sriram Shastry}}\ and\ \citenamefont
  {Sutherland}(1981)}]{shastry1981}%
  \BibitemOpen
  \bibfield  {author} {\bibinfo {author} {\bibfnamefont {B.}~\bibnamefont
  {{Sriram Shastry}}}\ and\ \bibinfo {author} {\bibfnamefont {B.}~\bibnamefont
  {Sutherland}},\ }\bibfield  {title} {\bibinfo {title} {Exact ground state of
  a quantum mechanical antiferromagnet},\ }\href
  {https://doi.org/https://doi.org/10.1016/0378-4363(81)90838-X} {\bibfield
  {journal} {\bibinfo  {journal} {Physica B+C}\ }\textbf {\bibinfo {volume}
  {108}},\ \bibinfo {pages} {1069} (\bibinfo {year} {1981})}\BibitemShut
  {NoStop}%
\bibitem [{\citenamefont {Miyahara}\ and\ \citenamefont
  {Ueda}(1999)}]{Miyahara1999}%
  \BibitemOpen
  \bibfield  {author} {\bibinfo {author} {\bibfnamefont {S.}~\bibnamefont
  {Miyahara}}\ and\ \bibinfo {author} {\bibfnamefont {K.}~\bibnamefont
  {Ueda}},\ }\bibfield  {title} {\bibinfo {title} {Exact dimer ground state of
  the two dimensional heisenberg spin system
  ${\mathrm{srcu}}_{2}({\mathrm{bo}}_{3}){}_{2}$},\ }\href
  {https://doi.org/10.1103/PhysRevLett.82.3701} {\bibfield  {journal} {\bibinfo
   {journal} {Phys. Rev. Lett.}\ }\textbf {\bibinfo {volume} {82}},\ \bibinfo
  {pages} {3701} (\bibinfo {year} {1999})}\BibitemShut {NoStop}%
\bibitem [{\citenamefont {Knetter}\ \emph {et~al.}(2000)\citenamefont
  {Knetter}, \citenamefont {B\"uhler}, \citenamefont {M\"uller-Hartmann},\ and\
  \citenamefont {Uhrig}}]{Knetter2000}%
  \BibitemOpen
  \bibfield  {author} {\bibinfo {author} {\bibfnamefont {C.}~\bibnamefont
  {Knetter}}, \bibinfo {author} {\bibfnamefont {A.}~\bibnamefont {B\"uhler}},
  \bibinfo {author} {\bibfnamefont {E.}~\bibnamefont {M\"uller-Hartmann}},\
  and\ \bibinfo {author} {\bibfnamefont {G.~S.}\ \bibnamefont {Uhrig}},\
  }\bibfield  {title} {\bibinfo {title} {Dispersion and symmetry of bound
  states in the shastry-sutherland model},\ }\href
  {https://doi.org/10.1103/PhysRevLett.85.3958} {\bibfield  {journal} {\bibinfo
   {journal} {Phys. Rev. Lett.}\ }\textbf {\bibinfo {volume} {85}},\ \bibinfo
  {pages} {3958} (\bibinfo {year} {2000})}\BibitemShut {NoStop}%
\bibitem [{\citenamefont {Totsuka}\ \emph {et~al.}(2001)\citenamefont
  {Totsuka}, \citenamefont {Miyahara},\ and\ \citenamefont
  {Ueda}}]{Totsuka2001}%
  \BibitemOpen
  \bibfield  {author} {\bibinfo {author} {\bibfnamefont {K.}~\bibnamefont
  {Totsuka}}, \bibinfo {author} {\bibfnamefont {S.}~\bibnamefont {Miyahara}},\
  and\ \bibinfo {author} {\bibfnamefont {K.}~\bibnamefont {Ueda}},\ }\bibfield
  {title} {\bibinfo {title} {Low-lying magnetic excitation of the
  shastry-sutherland model},\ }\href
  {https://doi.org/10.1103/PhysRevLett.86.520} {\bibfield  {journal} {\bibinfo
  {journal} {Phys. Rev. Lett.}\ }\textbf {\bibinfo {volume} {86}},\ \bibinfo
  {pages} {520} (\bibinfo {year} {2001})}\BibitemShut {NoStop}%
\bibitem [{\citenamefont {Matsubara}\ and\ \citenamefont
  {Matsuda}(1956)}]{MatsubaraMatsuda1956}%
  \BibitemOpen
  \bibfield  {author} {\bibinfo {author} {\bibfnamefont {T.}~\bibnamefont
  {Matsubara}}\ and\ \bibinfo {author} {\bibfnamefont {H.}~\bibnamefont
  {Matsuda}},\ }\bibfield  {title} {\bibinfo {title} {{A Lattice Model of
  Liquid Helium, I}},\ }\href {https://doi.org/10.1143/PTP.16.569} {\bibfield
  {journal} {\bibinfo  {journal} {Prog.\ Theor.\ Phys.}\ }\textbf {\bibinfo
  {volume} {16}},\ \bibinfo {pages} {569} (\bibinfo {year} {1956})}\BibitemShut
  {NoStop}%
\bibitem [{\citenamefont {Giamarchi}\ and\ \citenamefont
  {Tsvelik}(1999)}]{GiamarchiT}%
  \BibitemOpen
  \bibfield  {author} {\bibinfo {author} {\bibfnamefont {T.}~\bibnamefont
  {Giamarchi}}\ and\ \bibinfo {author} {\bibfnamefont {A.~M.}\ \bibnamefont
  {Tsvelik}},\ }\bibfield  {title} {\bibinfo {title} {Coupled ladders in a
  magnetic field},\ }\href {https://doi.org/10.1103/PhysRevB.59.11398}
  {\bibfield  {journal} {\bibinfo  {journal} {Phys. Rev. B}\ }\textbf {\bibinfo
  {volume} {59}},\ \bibinfo {pages} {11398} (\bibinfo {year}
  {1999})}\BibitemShut {NoStop}%
\bibitem [{\citenamefont {Misguich}\ and\ \citenamefont
  {Oshikawa}(2004)}]{MisguichO2004}%
  \BibitemOpen
  \bibfield  {author} {\bibinfo {author} {\bibfnamefont {G.}~\bibnamefont
  {Misguich}}\ and\ \bibinfo {author} {\bibfnamefont {M.}~\bibnamefont
  {Oshikawa}},\ }\bibfield  {title} {\bibinfo {title} {Bose–einstein
  condensation of magnons in tlcucl3: Phase diagram and specific heat from a
  self-consistent hartree–fock calculation with a realistic dispersion
  relation},\ }\href {https://doi.org/10.1143/JPSJ.73.3429} {\bibfield
  {journal} {\bibinfo  {journal} {J.\ Phys.\ Soc.\ Jpn.}\ }\textbf {\bibinfo
  {volume} {73}},\ \bibinfo {pages} {3429} (\bibinfo {year}
  {2004})}\BibitemShut {NoStop}%
\bibitem [{\citenamefont {Furukawa}\ and\ \citenamefont
  {Momoi}(2020)}]{FurukawaM2020}%
  \BibitemOpen
  \bibfield  {author} {\bibinfo {author} {\bibfnamefont {S.}~\bibnamefont
  {Furukawa}}\ and\ \bibinfo {author} {\bibfnamefont {T.}~\bibnamefont
  {Momoi}},\ }\bibfield  {title} {\bibinfo {title} {Effects of
  dzyaloshinskii–moriya interactions in volborthite: Magnetic orders and
  thermal hall effect},\ }\href {https://doi.org/10.7566/JPSJ.89.034711}
  {\bibfield  {journal} {\bibinfo  {journal} {J.\ Phys.\ Soc.\ Jpn.}\ }\textbf
  {\bibinfo {volume} {89}},\ \bibinfo {pages} {034711} (\bibinfo {year}
  {2020})}\BibitemShut {NoStop}%
\bibitem [{\citenamefont {Holstein}\ and\ \citenamefont
  {Primakoff}(1940)}]{HolsteinPrimakoff}%
  \BibitemOpen
  \bibfield  {author} {\bibinfo {author} {\bibfnamefont {T.}~\bibnamefont
  {Holstein}}\ and\ \bibinfo {author} {\bibfnamefont {H.}~\bibnamefont
  {Primakoff}},\ }\bibfield  {title} {\bibinfo {title} {Field dependence of the
  intrinsic domain magnetization of a ferromagnet},\ }\href
  {https://doi.org/10.1103/PhysRev.58.1098} {\bibfield  {journal} {\bibinfo
  {journal} {Phys. Rev.}\ }\textbf {\bibinfo {volume} {58}},\ \bibinfo {pages}
  {1098} (\bibinfo {year} {1940})}\BibitemShut {NoStop}%
\bibitem [{\citenamefont {Zhitomirsky}\ and\ \citenamefont
  {Chernyshev}(1999)}]{ZhitomirskyC}%
  \BibitemOpen
  \bibfield  {author} {\bibinfo {author} {\bibfnamefont {M.~E.}\ \bibnamefont
  {Zhitomirsky}}\ and\ \bibinfo {author} {\bibfnamefont {A.~L.}\ \bibnamefont
  {Chernyshev}},\ }\bibfield  {title} {\bibinfo {title} {Instability of
  antiferromagnetic magnons in strong fields},\ }\href
  {https://doi.org/10.1103/PhysRevLett.82.4536} {\bibfield  {journal} {\bibinfo
   {journal} {Phys. Rev. Lett.}\ }\textbf {\bibinfo {volume} {82}},\ \bibinfo
  {pages} {4536} (\bibinfo {year} {1999})}\BibitemShut {NoStop}%
\bibitem [{\citenamefont {Orignac}\ \emph {et~al.}(2007)\citenamefont
  {Orignac}, \citenamefont {Citro},\ and\ \citenamefont
  {Giamarchi}}]{OrignacCG}%
  \BibitemOpen
  \bibfield  {author} {\bibinfo {author} {\bibfnamefont {E.}~\bibnamefont
  {Orignac}}, \bibinfo {author} {\bibfnamefont {R.}~\bibnamefont {Citro}},\
  and\ \bibinfo {author} {\bibfnamefont {T.}~\bibnamefont {Giamarchi}},\
  }\bibfield  {title} {\bibinfo {title} {Critical properties and bose-einstein
  condensation in dimer spin systems},\ }\href
  {https://doi.org/10.1103/PhysRevB.75.140403} {\bibfield  {journal} {\bibinfo
  {journal} {Phys. Rev. B}\ }\textbf {\bibinfo {volume} {75}},\ \bibinfo
  {pages} {140403(R)} (\bibinfo {year} {2007})}\BibitemShut {NoStop}%
\bibitem [{\citenamefont {Starykh}\ and\ \citenamefont
  {Balents}(2014)}]{StarykhBalents2014}%
  \BibitemOpen
  \bibfield  {author} {\bibinfo {author} {\bibfnamefont {O.~A.}\ \bibnamefont
  {Starykh}}\ and\ \bibinfo {author} {\bibfnamefont {L.}~\bibnamefont
  {Balents}},\ }\bibfield  {title} {\bibinfo {title} {Excitations and
  quasi-one-dimensionality in field-induced nematic and spin density wave
  states},\ }\href {https://doi.org/10.1103/PhysRevB.89.104407} {\bibfield
  {journal} {\bibinfo  {journal} {Phys. Rev. B}\ }\textbf {\bibinfo {volume}
  {89}},\ \bibinfo {pages} {104407} (\bibinfo {year} {2014})}\BibitemShut
  {NoStop}%
\bibitem [{\citenamefont {Zhang}\ \emph {et~al.}(2017)\citenamefont {Zhang},
  \citenamefont {Kaushal}, \citenamefont {Dagotto},\ and\ \citenamefont
  {Batista}}]{ZhangBatista2017}%
  \BibitemOpen
  \bibfield  {author} {\bibinfo {author} {\bibfnamefont {S.-S.}\ \bibnamefont
  {Zhang}}, \bibinfo {author} {\bibfnamefont {N.}~\bibnamefont {Kaushal}},
  \bibinfo {author} {\bibfnamefont {E.}~\bibnamefont {Dagotto}},\ and\ \bibinfo
  {author} {\bibfnamefont {C.~D.}\ \bibnamefont {Batista}},\ }\bibfield
  {title} {\bibinfo {title} {Spin-orbit interaction driven dimerization in
  one-dimensional frustrated magnets},\ }\href
  {https://doi.org/10.1103/PhysRevB.96.214408} {\bibfield  {journal} {\bibinfo
  {journal} {Phys. Rev. B}\ }\textbf {\bibinfo {volume} {96}},\ \bibinfo
  {pages} {214408} (\bibinfo {year} {2017})}\BibitemShut {NoStop}%
\end{thebibliography}%

\end{document}